\documentclass[12pt,aps,tightenlines,notitlepage]{revtex4-1}
\usepackage{amsmath}
\usepackage{amssymb}
\usepackage{bm}
\usepackage[mathscr]{eucal}
\usepackage{theorem}
\usepackage{graphicx}
\usepackage{psfrag}
\usepackage{subfigure}
\usepackage{color}
\usepackage{wasysym}
\include{graphix}
\usepackage{framed}
\usepackage{enumitem}%
\usepackage{lipsum} 
\usepackage{float}

\newcommand{\ampl}{\epsilon}
\newcommand{\freq}{\omega}
\newcommand{\mygrav}{\mathcal{G}}
\newcommand{\mywe}{\mathrm{We}}
\newcommand{\myre}{\mathrm{Re}}
\newcommand{\mydrop}{\mathrm{d}p_0/\mathrm{d}x}
\newcommand{\myfs}{\xi}

\newcommand{\vecu}{\bm{u}}

\newcommand{\vecnhat}{\widehat{\bm{n}}}
\newcommand{\veczhat}{\widehat{\bm{z}}}

\newcommand{\wzonly}{\widetilde{w}}

\newcommand{\omegazonly}{\widetilde{\eta}}
\newcommand{\omegazonlyB}{\widetilde{\eta}_{B}}
\newcommand{\omegazonlyT}{\widetilde{\eta}_{T}}

\newcommand{\wzonlyab}{\widetilde{w}_{\alpha\beta}}
\newcommand{\omegazonlyab}{\widetilde{\eta}_{\alpha\beta}}

\graphicspath{{figs/}}

\newcommand{\qed}{\nobreak \ifvmode \relax \else
      \ifdim\lastskip<1.5em \hskip-\lastskip
      \hskip1.5em plus0em minus0.5em \fi \nobreak
      \vrule height0.75em width0.5em depth0.25em\fi}

\newcommand{\imag}{\mathrm{i}}
\newcommand{\mathe}{\mathrm{e}}
\newcommand{\mathd}{\mathrm{d}}

\newcommand{\omi}{\omega_{\mathrm{i}}}

\newcommand{\vecx}{\bm{x}}

\newcommand{\myell}{\mathcal{L}}
\newcommand{\myemm}{\mathcal{M}}
\newcommand{\mychi}{\bm{\chi}}

\newcommand{\myX}{\mathcal{E}}
\newcommand{\evolop}{\mathcal{E}}
\newcommand{\myG}{\mathbb{T}}

\newcommand{\vece}{\bm{e}}

\newcommand{\resmx}{\mathcal{R}(z)}

\begin{document}
%+Title
\title{Transient growth calculations obtained directly from the Orr--Sommerfeld matrices}
\author{Lennon \'O N\'araigh}
\affiliation{School of Mathematics and Statistics, University  College Dublin, Belfield, Dublin 4, Ireland}
%\affiliation{Complex and Adaptive Systems Laboratory, University  College Dublin, Belfield, Dublin 4, Ireland}

\date{\today}

\begin{abstract}
We introduce  and validate an algorithm to compute transient ampliﬁcation factors for the Orr--Sommerfeld--Squire linear theory for parallel two-phase flow.  We further introduce direct numerical simulation as a way of comparing the linear theory with early-stage wave growth in simulations.  The simulation results are drawn from a strongly supercritical parameter case wherein the modal growth rates are strong.  In this case, the modal growth dominates the transient growth.  
\end{abstract}

\maketitle

\section{Introduction}

Computing the optimal disturbance for the Orr--Sommerfeld--Squire linear theory is by now a standard numerical technique.  Here, we introduce our own in-house method to do this, and validate our efforts by comparing with well-established results from the literature.  We also recapitulate a method of direct numerical simulation to solve the two-phase Navier--Stokes equations beyond linear theory, into the domain of nonlinear waves and beyond.  In this way we can compare and contrast the linear theory with the (nonlinear) direct numerical simulations.  We thereby demonstrate effectively what is a null result, namely that in strongly supercritical paramter regimes where modal growth rates are strong, the effects of transient growth are negligible when compared to modal growth.
This work is to be read as supporting material, to support other more substantial works to appear elsewhere.  

\section{Mathematical Modelling}

\begin{figure}
	\centering
		\includegraphics[width=0.7\textwidth]{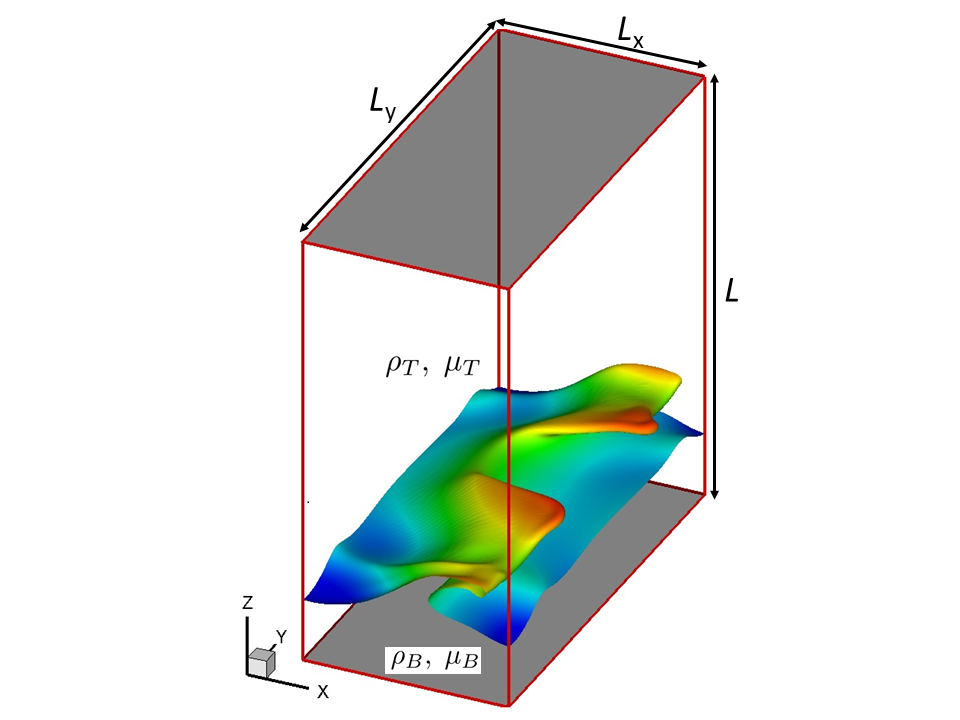}
		\caption{Definition sketch showing the three-dimensional problem geometry and the onset of interfacial waves.   The domain is periodic in the $x$- and $y$-directions.}
	\label{fig:schematic1}
\end{figure}

We study the system shown schematically in Figure~\ref{fig:schematic1}.
 A pressure drop drives the flow along the $x$-direction.  No-slip conditions are applied at $z=0$ and $z=1$.  The flow is initialized with a two-phase Poiseuille profile and a small-amplitude sinusoidal perturbation around an otherwise flat interface.  Waves develop as a result of linear instability and eventually overturn and form the complicated three-dimensional structure shown.

The two-phase Poiseuille flow profile and the flat interface correspond to an equilibrium base state, corresponding to a simple solution of the two-phase Navier--Stokes equations driven by a constant negative pressure drop $\mydrop$.   In this scenario, the Navier-Stokes equations  reduce to standard balances between pressure as well as viscous and gravitational forces.  This results in a flat interface $\myfs=0$, a unidirectional flow ($v=w=0$, and $u=U_0(z)$),  and a pressure field $p=(\mydrop)x$.
The analytic solution for the laminar velocity profile is then
\begin{equation}
U_0(z)=\begin{cases} U_B(z)=-\frac{\myre}{2m}z^2+Az,&0\leq z\leq  h_0,\\
                     U_T(z)=-\frac{\myre}{2}\left(z-1\right)^2+B\left(z-1\right),& h_0\leq z\leq 1\end{cases}.
\label{eq:basestate}
\end{equation}
The constants $A$ and $B$ are  determined from continuity of velocity and shear stress at the interface:
\begin{equation}
U_B(h_0)=U_T(h_0),\qquad m U_B'(h_0)=U_T'(h_0).
\end{equation}
In this way, the relevant length and velocity scales suggest themselves: the lengthscale is taken to be the channel height, and the standard velocity scale is taken as
\begin{equation}
V=\sqrt{-(L/\rho_T)(\mydrop)}.
\label{eq:vdef}
\end{equation}
Hence, the Reynolds number is $\myre=\rho_T VL/\mu_T$.  Of use later on will be the following furhter dimensionless groups:
\begin{equation}
\mygrav=\frac{gL}{V^2},\qquad 
\mywe=\frac{\rho_T LV^2}{\gamma}=\frac{L^2\left|\mydrop\right|}{\gamma},
\label{eq:scale1}
\end{equation}
where $g$ denotes the dimensional acceleration due to gravity and $\gamma$ is the surface tension.

\subsection{Linear Stability Analysis, including Transient Growth}

In the initial phase of the wave evolution in Figure~\ref{fig:schematic1}, the waves have a infinitesimally small amplitude.  As such, the corresponding Navier--Stokes equations in either phase (together with the interfacial matching conditions) can be linearized to yield Orr--Sommerfeld--Squire equations.  
As such, each flow variable is
expressed as a sum of the base state and the perturbation:
\begin{multline}
\myfs=h_0+\ampl\myfs_0\mathe^{\imag(\alpha x+\beta y-\freq t)},\qquad
w=\ampl\wzonly(z)\mathe^{\imag(\alpha x+\beta y-\freq t)},\qquad
\eta=\ampl\omegazonly(z)\mathe^{\imag(\alpha x+\beta y-\freq t)},\\
p=(\mydrop)x+\ampl\widetilde{p}(z)\mathe^{\imag(\alpha x+\beta y-\freq t)}.
\label{eq:pert1}
\end{multline}
Here $\epsilon$ is the infinitesimally small amplitude of the wave and $\myfs_0$ is its phase (with $|\myfs_0|=1$), and $\omega=c\alpha$, where $\omega$ is the complex frequency and $c$ the complex 
phase speed.
Substituting Equations~\eqref{eq:pert1} into the linearized Navier--Stokes equations (linearized around the base state~\eqref{eq:basestate}), we get the following
system of governing equations:
\begin{subequations}
\begin{eqnarray}
\imag\alpha r \myre\left[\left(\wzonly_B''-k^2\wzonly_B\right)\left(U_B-c\right)-\wzonly_BU_B''\right]&=&m\left(\wzonly_B''''-2k^2\wzonly_B''+k^4\wzonly_B\right),\\
\imag r \myre\left[\alpha\omegazonlyB\left(U_B-c\right)+\beta U_B'\wzonly_B\right]&=&m\left(\omegazonlyB''-k^2\omegazonlyB\right),
\end{eqnarray}
in the bottom phase, with $k^2=\alpha^2+\beta^2$, and
\begin{eqnarray}
\imag\alpha \myre\left[\left(\wzonly_T''-k^2\wzonly_T\right)\left(U_T-c\right)-\wzonly_TU_T''\right]&=&\wzonly_T''''-2k^2\wzonly_T''+k^4\wzonly_T,\\
\imag  \myre\left[\alpha\omegazonlyT\left(U_T-c\right)+\beta U_T'\wzonly_T\right]&=&\omegazonlyT''-k^2\omegazonlyT,
\end{eqnarray}%
\label{eq:osbulk}%
\end{subequations}%
in the top phase.  These are supplemented with the following no-slip and no-penetration boundary conditions:
\begin{equation}
\wzonly=\wzonly '=\omegazonly=0
\end{equation}
at the walls $z=0$ and $z=1$.  In addition,  matching conditions are prescribed at the interface $z=\myfs_0$.  In the streamwise direction, continuity of velocity and tangential stress and the jump condition in the normal stress imply the following relations: 
\begin{subequations}
\begin{eqnarray}
\wzonly_B&=&\wzonly_T,\\
\wzonly_B'+\myfs_0U_B'&=&\wzonly_T'+\myfs_0U_T',\qquad \myfs_0=\wzonly_B/(c-U_B)=\wzonly_T/(c-U_T),\\
m\left(\wzonly_B''+k^2\wzonly_B\right)&=&\wzonly_T''+k^2\wzonly_T,
\end{eqnarray}
\vspace{-.4in}
\begin{multline}
\imag \alpha r \myre\left[\wzonly_B\left(c-U_B\right)+\wzonly_B U_B'\right]+m\left(\wzonly_B'''-3k^2\wzonly_B\right)\\
=\imag \alpha \myre\left[\wzonly_T'\left(c-U_T\right)+\wzonly_TU'_T\right]+\left(\wzonly_T'''-3k^2\wzonly_T\right)
%
%=\frac{ k^4}{\mywe}
+k^2\left(\mygrav+\frac{k^2}{\mywe}\right)
\left[\frac{\wzonly_T'-\wzonly_B'}{\imag\alpha\left(U_B'-U_T'\right)}\right]=0.
\end{multline}
Finally, the same physical matching conditions applied to the spanwise direction give rise to the following relations:
\begin{eqnarray}
\omegazonlyB+\imag\beta U_B'\myfs_0&=&\omegazonlyT+\imag\beta U_T'\myfs_0,\\
m\omegazonlyB '&=&\omegazonlyT.
\end{eqnarray}%
\label{eq:ics_omegaz}%
\end{subequations}%
Equations~\eqref{eq:osbulk}--\eqref{eq:ics_omegaz} constitute an eigenvalue problem for the velocities $(\wzonly_B,\wzonly_T)$ and vorticity components $(\omegazonlyB,\omegazonlyT)$, with eigenvalue $\lambda=-\imag\alpha c=-\imag\omega$.  The equations can be further written in a generic operator/matirx form as follows:
in generic form for the eigenvalue $\lambda$:
\begin{equation}
\lambda
\left(\begin{array}{cc}\mathcal{M}_{OS} & 0\\
                       \mathcal{M}_{C} & \mathcal{M}_{S}\end{array}\right)\left(\begin{array}{c}\wzonlyab(z,\lambda)\\\omegazonlyab(z,\lambda)\end{array}\right)
=
\left(\begin{array}{cc}\mathcal{L}_{OS} & 0\\
                       \mathcal{L}_{C} & \mathcal{L}_{S}\end{array}\right)\left(\begin{array}{c}\wzonlyab(z,\lambda)\\\omegazonlyab(z,\lambda)\end{array}\right),
\label{eq:oss_laplace}
\end{equation}
The operator $\mathcal{L}_{OS}$ depends on wavenumbers and the wall-normal derivative, $\mathcal{L}_{OS}=\mathcal{L}_{OS}[\imag\alpha,\imag\beta,\partial_z]$ (similarly for the other operators). 
For wall-bounded flows, solution of Equation~\eqref{eq:oss_laplace} for the eigenvalue $\lambda$ gives a discrete family of eigenvalues for each Fourier mode, $\{\lambda_n(\alpha,\beta)\}_{n=0}^\infty$.  The growth rate is then determined by the eigenvalue with the largest real part: $\lambda=-\imag\omega$, with $\omi$ being the growth rate.
The eigenvalues can be computed approximately using numerical methods (e.g. Reference~\citep{Boomkamp1997}).

 The eigenvalue problem as described above in Equation~\eqref{eq:oss_laplace} can be viewed as the Laplace-transform of a differential-algebraic equation, the asymptotic solution of which  picks out the most-dangerous eigenmode of Equation~\eqref{eq:oss_laplace}.  However, at finite times, a combination of modes can combine to produce transient growth rates in excess of the asymptotic most-dangerous exponential growth rate.  This is possible  because the eigenfunctions of Equation~\eqref{eq:oss_laplace} are non-orthogonal, which is due in turn to the non-normality of the operators in the same equation~\citep{Trefethan1993,Schmid2001}.
The resulting transient growth is captured by the maximum amplification factor at a particular wavenumber ($\alpha , \beta $)
\begin{equation}
G_{\alpha\beta}(t)=\sup \|\left(\wzonlyab(z,t),\omegazonlyab(z,t)\right)\|_E,
\label{eq:galphabeta}
\end{equation}
where $\|\cdot\|_E$ denotes the energy norm~\citep{Schmid2001,Boomkamp1998} and the supremum in Equation~\eqref{eq:galphabeta} is taken over all possible states  whose energy norm at $t=0$ is unity.  For completeness, the energy norm is restated here: let $\wzonly_T$ and $\wzonly_B$ denote the wall-normal velocity $\wzonlyab$ in the top and bottom layers, and similarly for the wall-normal vorticity.  Then
\begin{multline}
\|\left(\wzonlyab(z,t),\omegazonlyab(z,t)\right)\|_E
=\frac{r}{2k^2}\int_{0}^h \left[\left(\frac{\mathd w_L}{\mathd z}\right)^2+k^2|w_L|^2+|\eta_L|^2\right]\mathd z\\
+
\frac{1}{2k^2}\int_{h}^1 \left[\left(\frac{\mathd w_G}{\mathd z}\right)^2+k^2|w_G|^2+|\eta_G|^2\right]\mathd z
+\tfrac{1}{2}\left[(r-1)\mygrav+\frac{k^2}{\mywe}\right]|\myfs|^2,\qquad k^2=\alpha^2+\beta^2,
\label{eq:enormdef}
\end{multline}
where $h$ is the dimensionless height of the lower layer, and $\myfs$ is the disturbed height of the interface relative to $h$.
Conceptually, the computation of the maximum amplification factor is achieved as follows.  Consideration is given to an arbitrary initial condition made up of a superposition of Orr--Sommerfeld--Squire modes at a fixed wavenumber $(\alpha,\beta)$.  The weights in this superposition are normalized such that the initial condition has unit energy norm.  The initial condition is then evolved forward to some later time $t$ under the linearized dynamics and the energy norm at a later time is checked.  The weights in the initial condition are then adjusted (subject to the unit-norm condition being  maintained) so that the energy norm at time $t$ attains its maximum possible value, which is precisely $G_{\alpha\beta}(t)$ in Equation~\eqref{eq:galphabeta}.
In practice, computation of the maximum amplification factor is standard but is nevertheless described in detail below in Section~\ref{sec:background}.  A connection between the transient characterization~\eqref{eq:galphabeta} and the eigenvalue analysis is obtained via the relation
\begin{equation}
\omi(\alpha,\beta)=\lim_{\substack{t_1,t_2\rightarrow\infty\\t_2\gg t_1}}\frac{1}{t_2-t_1}\log\left(\frac{G_{\alpha\beta}(t_2)}{G_{\alpha\beta}(t_1)}\right).
\end{equation}

\subsection{Direct Numerical Simulation}

Beyond linear theory, a levelset method with a continuous surface tension model~\citep{Sussman1999} is used as a model for the two-phase Navier--Stokes equations with the physical matching conditions at the interface:
\begin{subequations}
\begin{equation}
\rho(\phi)\left(\frac{\partial \bm{u}}{\partial t}+\bm{u}\cdot\nabla\bm{u}\right)=-\nabla p+\frac{1}{\myre}\nabla\cdot\left[\mu\left(\nabla \bm{u}+\nabla\bm{u}^T\right)\right]
+\frac{1}{\mywe}\delta_\epsilon(\phi)\kappa \vecnhat -\mygrav\rho(\phi)\veczhat,
%-\frac{1}{\mywe}\delta_\epsilon(\phi)\vecnhat\nabla\cdot\vecnhat-\mygrav\rho(\phi)\veczhat,
%-\rho \grav \mathbf{e}_z,
\label{eq:ns_def_mom}
\end{equation}%
\begin{equation}
\nabla\cdot\bm{u}=0,
\end{equation}%
\begin{equation}
\vecnhat=\frac{\nabla\phi}{|\nabla\phi|},\qquad \kappa=-\nabla\cdot\vecnhat,\qquad \frac{\partial \phi}{\partial t}+\bm{u}\cdot\nabla\phi=0,
\label{eq:ls_def}
\end{equation}%
\label{eq:ns_def}%
\end{subequations}%
%
%\rho&=&r\left(1-H(\phi)\right)+H(\phi),\\
%\mu&=&m\left(1-H(\phi)\right)+H(\phi),\\
%
where $\veczhat$ is the unit vector in the $z$-direction. 
Here also, $\phi(\vecx,t)$ is the levelset function indicating in which phase the point $\vecx$ lies ($\phi<0$ in the bottom layer, $\phi>0$ in the top layer).  The (possibly multivalued) interface $\myfs(\vecx,t)$  is therefore the zero level set, $\phi(\vecx,t)=0\implies \vecx=(x,y,\myfs(x,y,t))$.
Moreover, the levelset function determines the unit vector normal to the interface $\vecnhat$, as well as the density and viscosity, via the relations
$\rho=r\left(1-H_\epsilon(\phi)\right)+H_\epsilon(\phi)$ and 
$\mu=m\left(1-H_\epsilon(\phi)\right)+H_\epsilon(\phi)$ respectively.  The function $H_\epsilon(\phi)$ is a regularized Heaviside function, which is smooth across a width $\epsilon=1.5\Delta x$.  Finally, $\delta_\epsilon(s)=dH_\epsilon(s)/ds$ is a regularized delta function supported on an interval $[-\epsilon,\epsilon]$.

Equations~\eqref{eq:ns_def} are solved in a density-contrast implementation of the computational framework TPLS~\citep{Naraigh2014linear,tpls_code}.   Specifically, the equations are discretized in space using an isotropic MAC grid wherein vector quantities are defined at cell faces and scalar quantities at the respective cell centres.  In terms of the temporal discretization, a third-order Adams--Bashforth scheme is used to treat the convective derivative, while the momentum fluxes are treated using the Crank--Nicolson method.  Pressure and the associated incompressibility of the flow are treated using a standard projection  method~\citep{chorin1968numerical}.  
The levelset function is advected using a third-order (fifth-order accurate) WENO scheme~\citep{jiang1996efficient}, which  is subsequently reinitialized using a Hamilton--Jacobi equation. 

Validation tests have  been performed on the code  wherein it is shown that the numerical simulations agree excellently with linear Orr--Sommerfeld--Squire theory.  Further validation of the code particular to ligament formation in density-contrasted flows has been carried out with respect to benchmarks  in the literature, 
where the code again shows excellent agreement with established numerical test cases. Further details of these tests are in References~\cite{Naraigh2014linear} and~\cite{solomenko2016}.

\section{Transient Growth -- detailed description of the methodology}
\label{sec:background}

We develop a numerical technique to compute the maximum norm in Equation~\eqref{eq:enormdef}.  The approach in this section is to build the method from the very beginning.  As such, we begin with a preliminary discussion focused on single-phase flows, wherein we can compare our computations for the amplification factor with well-established results in the literature.  Thereafter, we extend our method to two-phase flows.

\subsection{Preliminary Discussion}

The Orr--Sommerfeld--Squire equation for a generic single-phase problem in the hydrodynamic stability of a parallel flow can be written down in generic form as follows:
\begin{equation}
\myell \mychi=\lambda \myemm\mychi,
\label{eq:os_eig}
\end{equation}
The stability problem is solved at a particular set of wavenumbers $(\alpha,\beta)$, and the Orr--Sommerfeld--Squire matrices $\myell$ and $\myemm$ and the eigenvalue $\lambda$ all depend on the wavenumbers.  The extension to the two-phase case is carried out below in Section~\ref{sec:twophase}.  Recall, the state vector $\mychi$ is obtained by writing the wall-normal velocity and vorticity in a finite Chebyshev approximation:
\[
w(z)=\sum_{i=0}^n A_i T_i(x),\qquad \eta=\sum_{i=0}^n B_i T_i(x),\qquad x=2z-1,
\]
such that
\[
\mychi=(A_0,\cdots,A_n,B_0,\cdots,B_n)^T.
\]

In our previous work~\cite{Onaraigh2013b}, we have demonstrated how the matrices in Equation~\eqref{eq:os_eig} can be used to solve the corresponding initial-value problem.  The initial-value problem is formulated as follows:
\begin{subequations}
\begin{equation}
\frac{\partial}{\partial t}\myemm\mychi=\myell \mychi,\qquad t>0,
\label{eq:os_ivp}
\end{equation}
with initial condition
\begin{equation}
\mychi(t=0)=\mychi_0,\qquad \mychi_0=(x_0,\cdots,x_n,y_0,\cdots,y_n)^T,
\label{eq:os_ivp_init}
\end{equation}%
\label{eq:os_ivp_all}%
\end{subequations}%
and where
\begin{subequations}
\begin{eqnarray}
x_0&=&\frac{1}{\pi}\int_{-1}^1 T_0(x)w\left(\tfrac{1}{2}(x+1),t=0\right)\frac{\mathd x}{\sqrt{1-x^2}},\\
x_i&=&\frac{2}{\pi}\int_{-1}^1 T_i(x)w\left(\tfrac{1}{2}(x+1),t=0\right)\frac{\mathd x}{\sqrt{1-x^2}},\qquad i\neq 0,\\
y_0&=&\frac{1}{\pi}\int_{-1}^1 T_0(x)\eta\left(\tfrac{1}{2}(x+1),t=0\right)\frac{\mathd x}{\sqrt{1-x^2}},\\
y_i&=&\frac{2}{\pi}\int_{-1}^1 T_i(x)\eta\left(\tfrac{1}{2}(x+1),t=0\right)\frac{\mathd x}{\sqrt{1-x^2}},\qquad i\neq 0.
\end{eqnarray}
\end{subequations}%

\subsection*{The evolution operator}

In Reference~\cite{Onaraigh2013b}, we have shown how the solution to Equation~\eqref{eq:os_ivp_all} can be written as
\[
\mychi(t_p)=\evolop_{t_p}\mychi_0,\qquad t_p=p\Delta t,\qquad p=0,1,\cdots,
\]
where $\Delta t\rightarrow 0$, keeping $t_p=t$ finite.  Also $\evolop_t$ is the {\textbf{evolution operator}}, where
\begin{equation}
\evolop_t=\lim_{\Delta t\rightarrow 0}\left[\left(\myemm-\Delta t\myell\right)^{-1}\myemm\right]^p.
\label{eq:backward}
\end{equation}
Note that Equation~\eqref{eq:backward} amounts to solving the linear differential algebraic equation  (DAE)~\eqref{eq:os_ivp} using the backward Euler method.  Previously (e.g. in Reference~\cite{Onaraigh2013b}) we used a trapezoidal rule.  However, in the present calculations, we have found by trial and error that the backward Euler method is the most accurate one for our purposes, and a particular advantage of the Backward Euler method is its large domain of stability.

\subsection*{Alternative derivation of the evolution operator}

An obvious solution to Equation~\eqref{eq:os_ivp_all} (yet equivalent to the one in Equation~\eqref{eq:backward})  is
\begin{subequations}
\begin{equation}
\mychi(t)=\sum_{q=1}^N\mu_q\mathe^{\lambda_qt}\mychi_{(q)},
\label{eq:os_sln_eigs1}%
\end{equation}
where $(\mychi_{(q)},\lambda_q)$ are an eigenvector-eigenvalue pair:
\begin{equation}
\mathcal{L}\mychi_{(q)}=\lambda_{(q)}\mathcal{M}\mychi_{(q)},\qquad q=1,2,\cdots,N,
\end{equation}
where $N=2(n+1)$ for the problem in Equation~\eqref{eq:os_ivp_all}.
This solution is a complete solution provided the eigenvectors form a complete basis, that is, the matrix
\begin{equation}
V=\left(\begin{array}{ccc} | &  & |\\
                           \mychi_{(1)}&\cdots&\mychi_{(N)}\\
													 | &  & |\\
 \end{array}\right),
\end{equation}
\label{eq:os_sln_eigs}%
\end{subequations}%
is invertible.  We work with the assumption that $V$ is indeed invertible, but that $V^\dagger V$ is not a diagonal matrix because the eigenvalue problem is non-normal.  The problem here is to relate the $\mu_q$-coefficients to the initial data in Equation~\eqref{eq:os_ivp_init}.  

We define
\[
\chi_{i0}=\langle\vece_i,\mychi_0\rangle,\qquad \chi_i(t)=\langle\vece_i,\mychi(t)\rangle
\]
where $\vece_i$ denotes the usual basis in $\mathbb{C}^N$, $\langle\cdot,\rangle$ denotes the usual scalar product on the same, and
$\mychi_0=(\chi_{01},\cdots,\chi_{0N})^T$.  From Equation~\eqref{eq:os_sln_eigs1} we have
\begin{eqnarray*}
\chi_{i0}&=&\langle\vece_i,\mychi_0\rangle,\\
&=&\sum_{q}\mu_q \langle\vece_i,\mychi_{(q)}\rangle,\\
&=&\sum_{q}\mu_q V_{qi},
\end{eqnarray*}
hence
\[
(\mu_1,\cdots,\mu_N)^T=V^{-1}\mychi_0.
\]
Also from Equation~\eqref{eq:os_sln_eigs1} we have
\begin{eqnarray*}
\chi_{i}(t)&=&\langle\vece_i,\mychi(t)\rangle,\\
&=&\sum_{q}\mu_q \mathe^{\lambda_q t} \langle\vece_i,\mychi_{(q)}\rangle,\\
&=&\sum_{q}\mu_q \mathe^{\lambda_q t} V_{iq},\\
&=&\sum_{q}\left(V^{-1}\mychi_0\right)_q \mathe^{\lambda_q t} V_{iq},\\
&=&\sum_{q}\sum_{j} V_{iq}\mathe^{\lambda_q t}V^{-1}_{qj}\chi_{0j},\\
&=&\sum_j \left(V\mathe^{\Lambda t}V^{-1}\right)_{ij}\chi_{0j},
\end{eqnarray*}
hence
\[
\mychi(t)=\left(V\mathe^{\Lambda t}V^{-1}\right)\mychi_0,
\]
and
\begin{equation}
\evolop_t=V\mathe^{\Lambda t}V^{-1},
\label{eq:evolop_eigs}
\end{equation}
where
\[
\mathe^{\Lambda t}=\mathrm{diag}\left(\mathe^{\lambda_{(1)}t},\cdots,\mathe^{\lambda_{(N)}t}\right),
\]
It therefore follows from these calculations and from Equation~\eqref{eq:backward} that
\begin{equation}
\evolop_t=\lim_{\Delta t\rightarrow 0}\left[\left(\myemm-\Delta t\myell\right)^{-1}\myemm\right]^p
=V\mathe^{\Lambda t}V^{-1}
\label{eq:evolop_all}
\end{equation}
where again the limit $\Delta t\rightarrow 0$ is taken in Equation~\eqref{eq:evolop_all}, keeping $t_p=p\Delta t=t$ finite.  Throughout this work, both methods have been used with identical results, although using the eigenvalue/eigenvector method has proved more expedient in certain situations.

\subsection{The basic method}
\label{sec:basic}

The idea of the transient-growth calculation is to start with the energy norm
\begin{equation}
E(t)=\frac{1}{2k^2}\int_0^1 \left(|\partial_z w|^2+k^2|w|^2+|\eta|^2\right)\mathd z,\qquad k^2=\alpha^2+\beta^2,
\label{eq:et}
\end{equation}
and at each point in time to optimize the energy norm subject to the constraint that $E(0)=1$.  The resulting maximum energy is called the \textit{maximum amplification factor}, $G(t)$.  These calculations can be done within the framework of Section~\ref{sec:background} as follows.  First, the energy norm in Equation~\eqref{eq:et} is identified with a scalar product on the space of admissible $\mychi$-vectors:
\begin{eqnarray*}
E(t)&=&\tfrac{1}{2}\frac{1}{2k^2}\int_{-1}^1 \mathd x\left[\left|\sum_{i=0}^n A_i  e_x T_i'(x)\right|^2+k^2\left|\sum_{i=0}^n A_i T_i(x)\right|^2+\left|\sum_{i=0}^n B_i T_i(x)\right|^2\right],\qquad e_x=\frac{dx}{dz}=2,\\
&=&\frac{1}{2k^2}\sum_{i,j}A_i^* A_j  \left(\tfrac{1}{2}e_x^2\int_{-1}^1 T_i'(x)T_j'(x)\mathd x\right)
+\frac{1}{2k^2}\sum_{i,j}A_i^* A_j \left(\tfrac{1}{2}k^2\int_{-1}^1 T_i(x)T_j(x)\mathd x\right)\\
&\phantom{=}&\phantom{aaaaa}+\frac{1}{2k^2}\sum_{i,j}B_i^* B_j \left(\tfrac{1}{2}\int_{-1}^1 T_i(x)T_j(x)\mathd x\right).
\end{eqnarray*}
Call
\begin{eqnarray*}
\mathbb{T}^{(1)}_{ij}&=&\tfrac{1}{2}e_x^2\int_{-1}^1 T_i'(x)T_j'(x)\mathd x,\\
\mathbb{T}^{(0)}_{ij}&=&\tfrac{1}{2}\int_{-1}^1 T_i(x)T_j(x)\mathd x.
\end{eqnarray*}
We have
\begin{eqnarray*}
E(t)&=& \frac{1}{2k^2}\left[\sum_{i,j}A_i^* A_j \mathbb{T}^{(1)}_{ij}+ k^2\sum_{i,j}A_i^* A_j \mathbb{T}^{(0)}_{ij}+ \sum_{i,j}B_i^* B_j \mathbb{T}^{(0)}_{ij}\right],\\
&=&\frac{1}{2k^2}\bigg\langle \mychi,\left(\begin{array}{cc} \mathbb{T}^{(1)} + k^2 \mathbb{T}^{(0)}& 0\\ 0 & \mathbb{T}^{(0)}\end{array}\right)\mychi\bigg\rangle,\\
&:=&\frac{1}{2k^2}\langle \mychi,\myG\mychi\rangle.
\end{eqnarray*}
Here, the angle brackets denote the usual scalar product  on the space of $\mychi$-vectors, and the matrix $\myG$ is symmetric positive-definite. Thus, the equation
\[
E(t)=\frac{1}{2k^2}\langle \mychi,\myG\mychi\rangle
\]
defines a scalar product on the space of $\mychi$-vectors.  
However, we have
\[
\mychi=\myX_t\mychi_0,
\]
hence
\[
E(t)=\frac{1}{2k^2}\langle \mychi_0 ,\myX_t^\dagger\myG\myX_t\mychi_0\rangle.
\]
Thus, the optimization to be performed can be recast as an optimization of the functional
\[
E[\mychi_0]=\frac{1}{2k^2}\langle \mychi_0, \myX_t^\dagger\myG\myX_t\mychi_0\rangle,
\]
subject to the constraint that
\[
\frac{1}{2k^2}\langle \mychi_0 ,\myG \mychi_0\rangle=1.
\]
In other words, we have the following Lagrange-multiplier problem:
\[
E[\mychi_0]=\frac{1}{2k^2}\langle \mychi_0, \myX_t^\dagger\myG\myX_t\mychi_0\rangle-\lambda\left(\frac{1}{2k^2}\langle \mychi_0, \myG\mychi_0\rangle-1\right).
\]
The optimum vector is obtained by setting
\[
\frac{\delta E}{\delta\mychi_0^*}=0,
\]
in other words,
\begin{equation}
\myX^\dagger\myG\myX_t\mychi_0=\lambda  \myG\mychi_0.
\label{eq:eig_opt}
\end{equation}
Equation~\eqref{eq:eig_opt} is a generalized eigenvalue problem, and it is readily checked that the eigenvalues are real (both matrices appearing in the problem are Hermitian) and moreover, that the eigenvalues are non-negative.  It can be further shown by a straightforward calculation (backsubstitution into the constrained functional) that 
\[
\sup_{\mychi_0} \left[E[\mychi_0]-\lambda\left(\frac{1}{2k^2}\langle \mychi_0, \myG\mychi_0\rangle-1\right)\right]=\max \lambda,
\]
where the maximum is taken over the spectrum of the generalized eigenvalue problem~\eqref{eq:eig_opt}.  Thus, at each point in time, the maximum amplification factor is
\[
G(t)=\max \lambda.
\]

\subsection{Validation -- single-phase flow}

\begin{figure}[htb]
	\centering
		\subfigure[]{\includegraphics[width=0.45\textwidth]{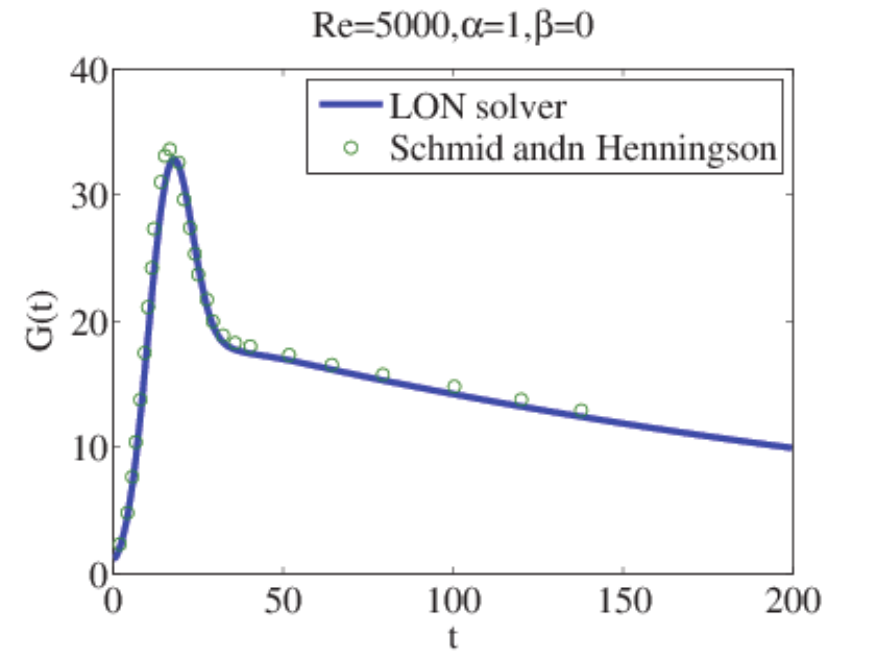}}
		\subfigure[]{\includegraphics[width=0.45\textwidth]{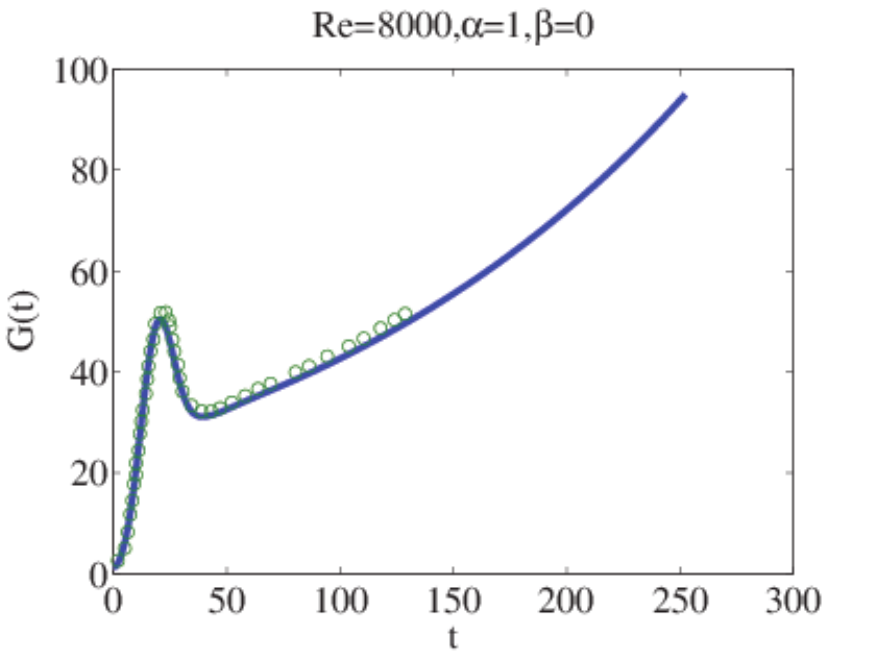}}
		\caption{Validation of our code for the  maximum amplification factor compared to known benchmark case in the literature (data from Reference~\cite{Schmid2001}).  The small discrepancies between the two datasets are due to errors in scanning and digitizing the data from the reference text.}
	\label{fig:validate}
\end{figure}

We have validated this procedure against the known test case of Poiseuille flow.  We work in the units used by Orzag and other later researchers for their stability calculations of single-phase Poiseuille flow~\cite{orzag1971}.  Thus, we take $\alpha=1$, $\beta=0$, and two cases for the Reynolds number: $Re=5000$ (asymptotically stable) and $Re=8000$ (asymptotically unstable).  A comparison between known results for $G(t)$ in this instance and the results from our own calculations is shown in Figure~\ref{fig:validate}.

\begin{figure}[htb]
	\centering
		\subfigure[$\,\,t=0.05$]{\includegraphics[width=0.32\textwidth]{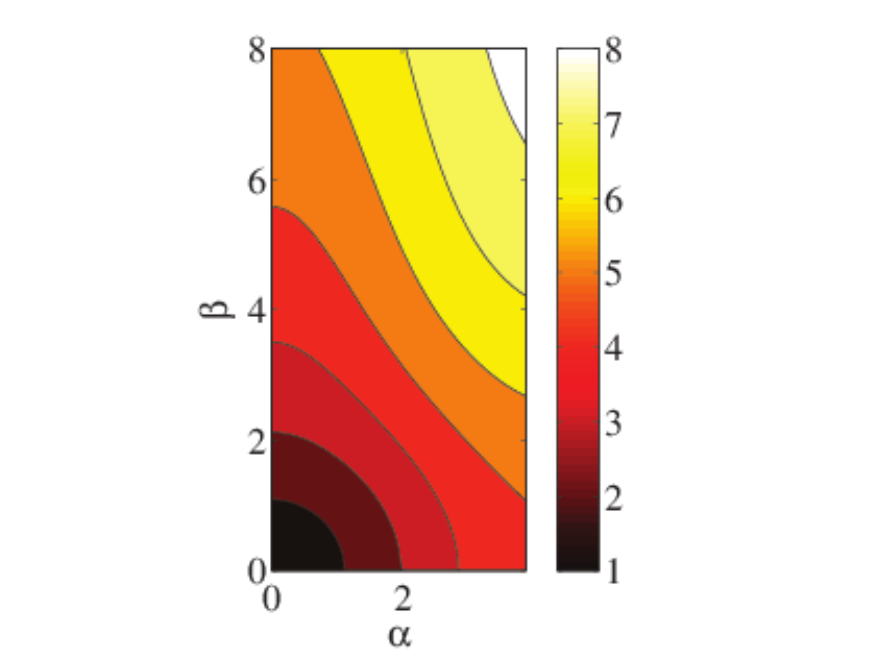}}
		\subfigure[$\,\,t=0.1$]{\includegraphics[width=0.32\textwidth]{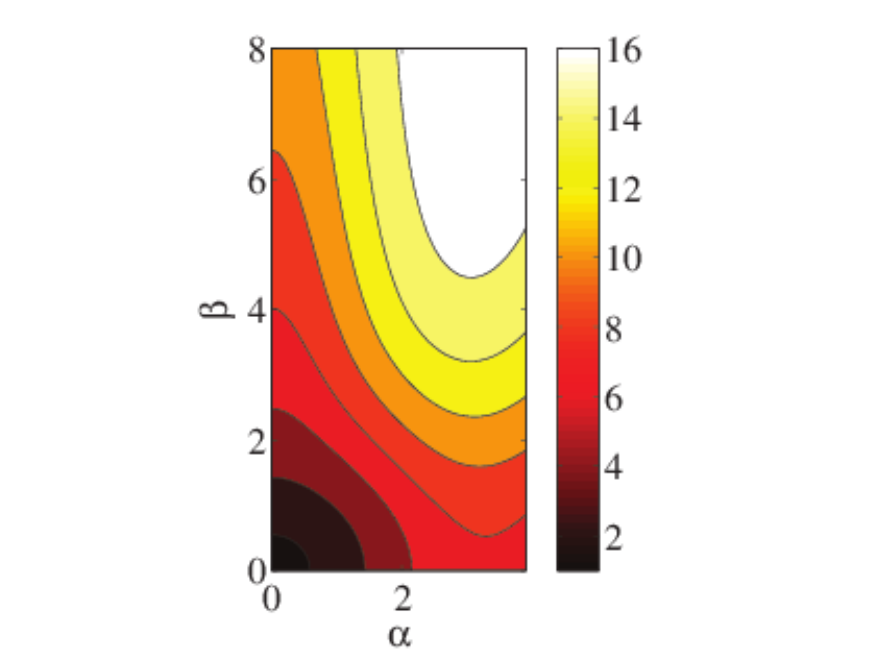}}
		\subfigure[$\,\,t=0.15$]{\includegraphics[width=0.32\textwidth]{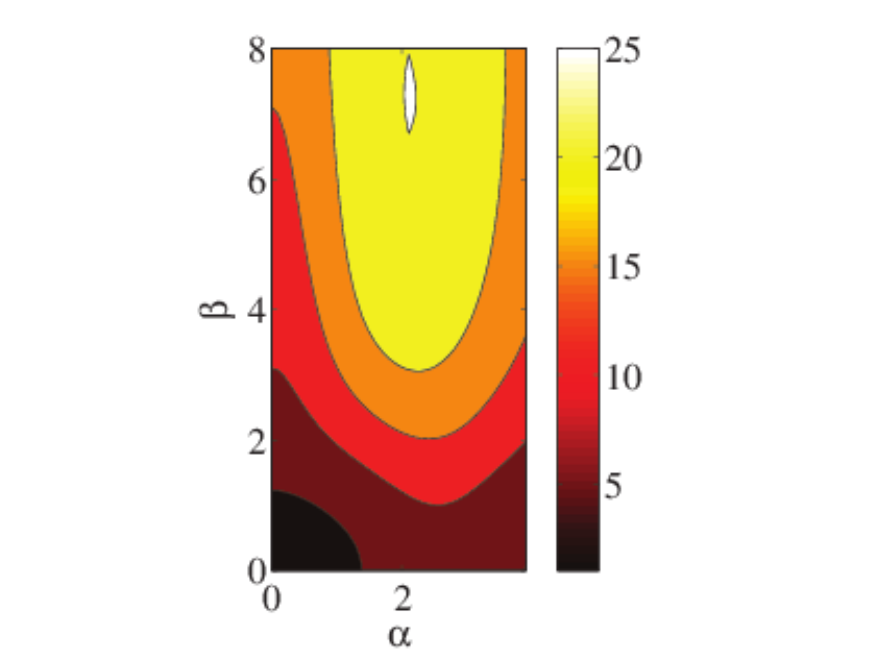}}
		\subfigure[$\,\,t=0.5$]{\includegraphics[width=0.32\textwidth]{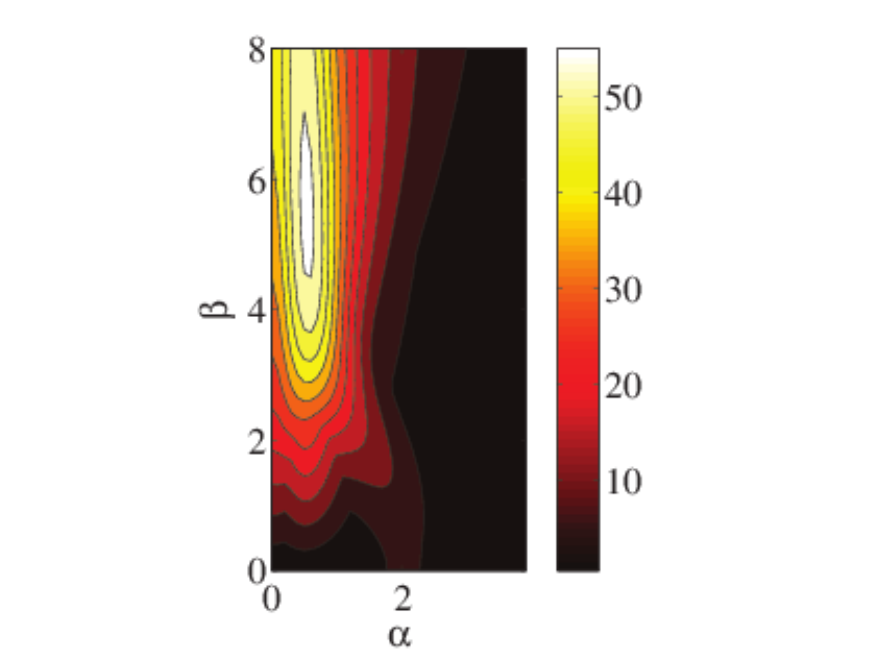}}
		\subfigure[$\,\,t=1$]{\includegraphics[width=0.32\textwidth]{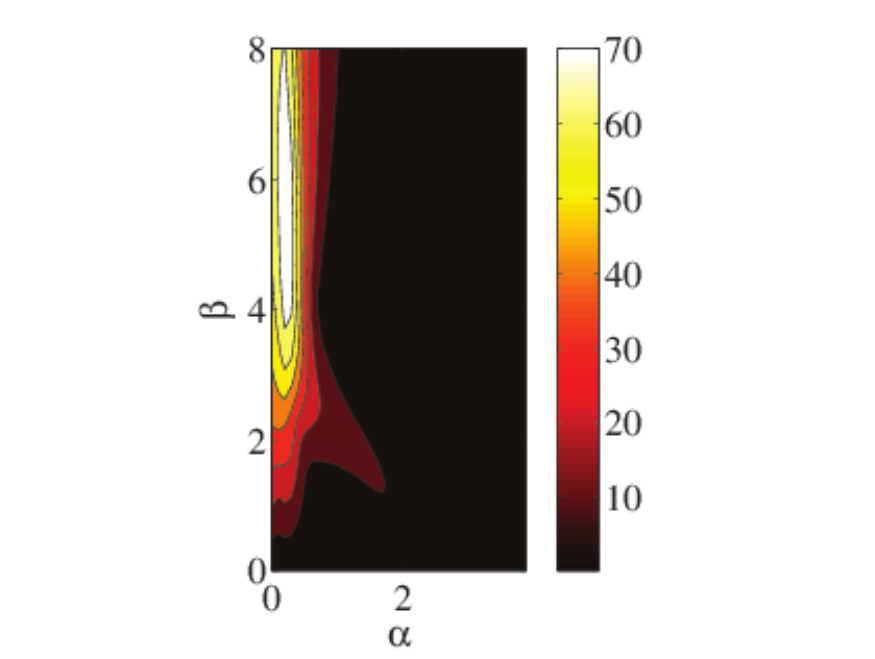}}
		\subfigure[$\,\,t=2$]{\includegraphics[width=0.32\textwidth]{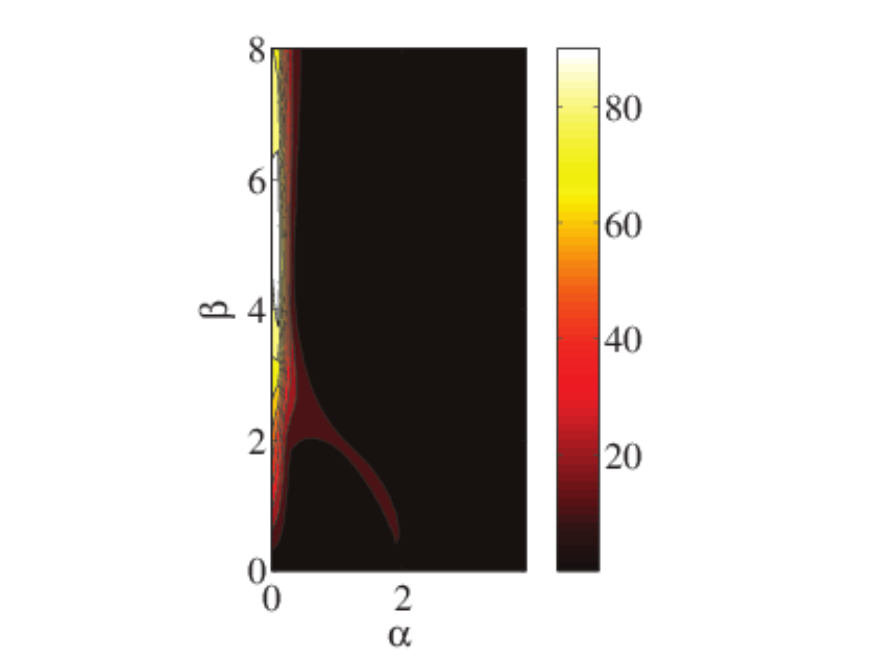}}
		\subfigure[$\,\,t=5$]{\includegraphics[width=0.32\textwidth]{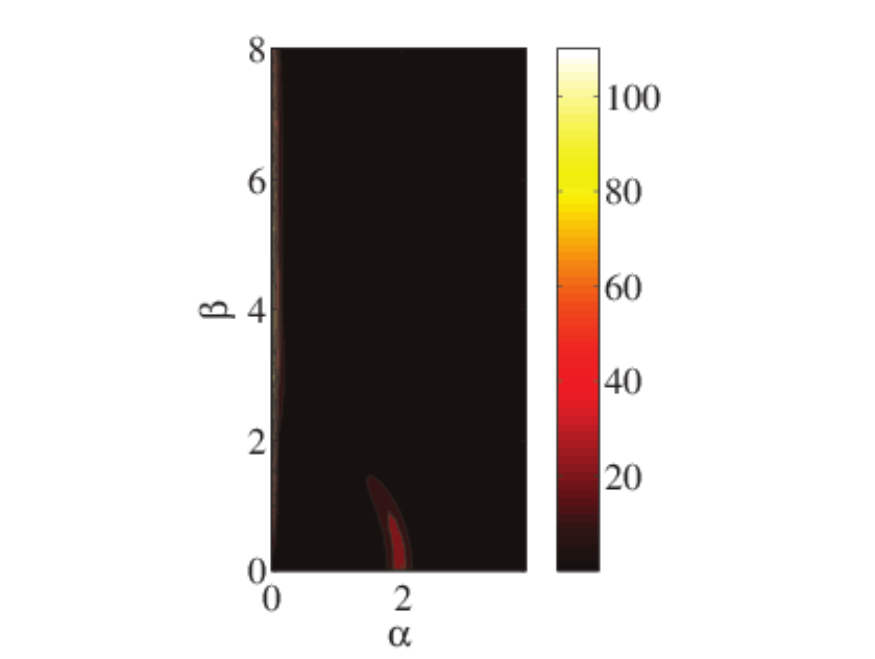}}
		\subfigure[$\,\,t=10$]{\includegraphics[width=0.32\textwidth]{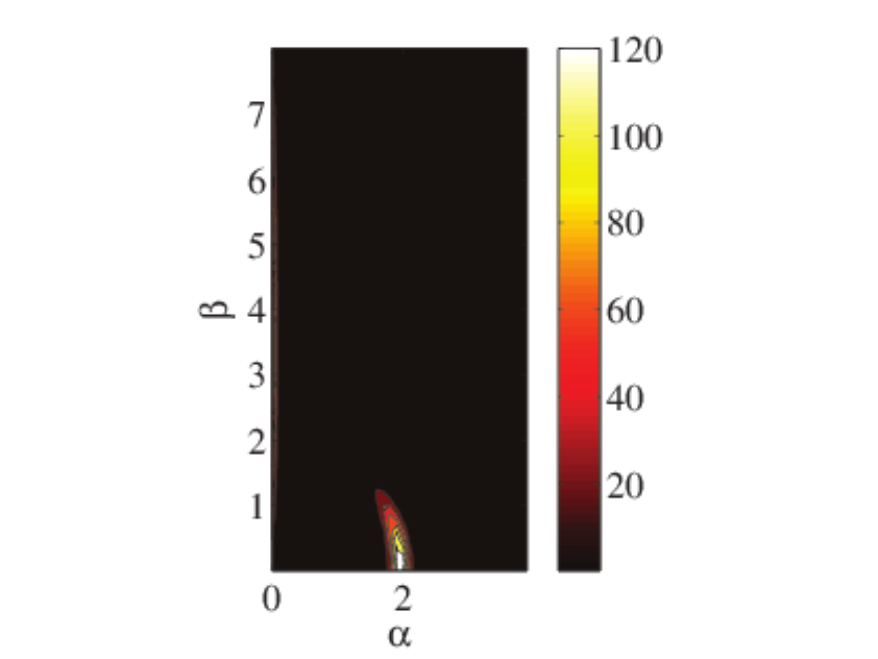}}
		\caption{Time evolution of the maximum amplification factor as a function of the wavenumbers $\alpha$ (streamwise) and $\beta$ (spanwise).  Between $t=0.1$ and $t=10$ the optimal disturbance moves from being spanwise-dominated to streamwise-dominated.}
	\label{fig:alphabeta}
\end{figure}
\begin{figure}
	\centering
		\includegraphics[width=0.6\textwidth]{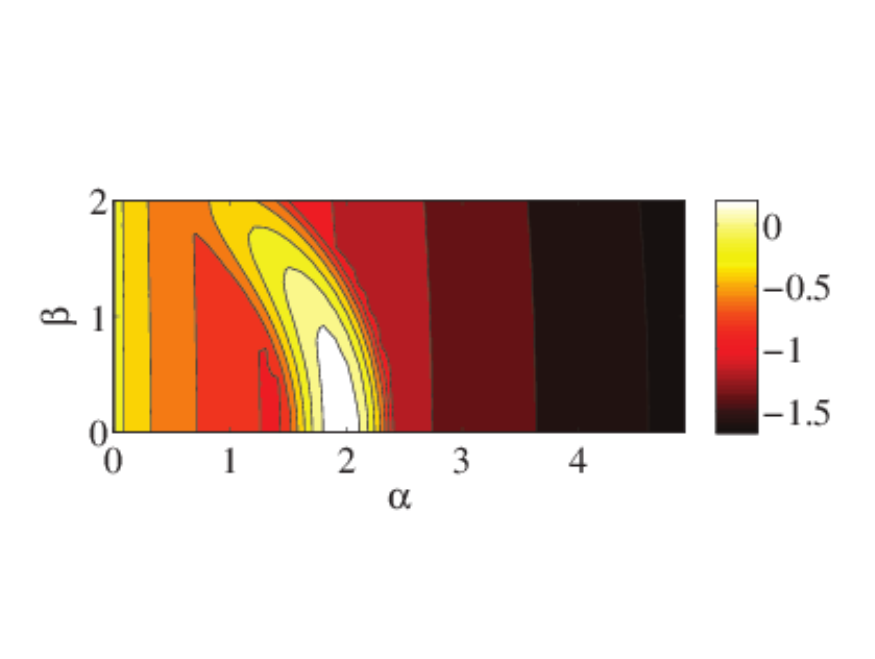}
		\caption{Single-phase flow eigenvalue analysis, corresponding to $t\rightarrow\infty$ and to be read in conjunction with Figure~\ref{fig:alphabeta}: Eigenvalue of most-dangerous mode of the Orr--Sommerfeld--Squire equations, with $Re=5000$.  The most-dangerous mode according to eigenvalue analysis (valid as $t\rightarrow\infty$) is a streamwise one.}
	\label{fig:eigs}
\end{figure}

It is now of interest to examine the behavior demonstrated in Figure~\ref{fig:validate} a little further.  We go back over to our own units based on the full channel height and the friction velocity and examine the features of the transient growth in the supercritical case $Re=8000$, $Re_*=\sqrt{8\times 8000}\approx  252.9822$, for various times $t\in [0,2]$ (corresponding to times $[0,2]Re_*/2$ in Figure~\ref{fig:validate}).  The resulting study is presented in Figure~\ref{fig:validate} where it should be noticed that it is the square root of the energy of the most-amplified disturbance that is plotted in a wavenumber space, for different $t$-values.
For very short times ($t=0.1$) the transiently most-amplified mode has a wavevector with components in both the streamwise and spanwise directions (at $t=0.1$, $\max_{\alpha,\beta}G_{\alpha,\beta}(t)$ occurs at $(\alpha,\beta)\approx (3,8)$).  As time goes by, the most-amplified mode moves to a more spanwise wavenumber such that by $t=1$ the maximum value $\max_{\alpha,\beta}G_{\alpha,\beta}(t)$ occurs at $\alpha\approx 0$ and $\beta=6$.  Thereafter, there is a slow evolution of the trajectory of the most-amplified disturbance through the wavenumber space away from spanwise wavenumbers towards streamwise ones (the eigenvalue theory predicts that as $t\rightarrow\infty$ the most-amplified disturbance is a streamwise-only mode -- Figure~\ref{fig:eigs}).  By $t=10$ the asymptotic state is reached and the most-amplified disturbance is indeed streamwise-only.

\subsection{Two-phase flows}
\label{sec:twophase}

Application of the basic method described above to two-phase flows with surface tension has produced significant quantitative differences between the present method and the existing test cases in the literature (e.g. Reference~\cite{yecko2008disturbance}), albeit that the qualitative trends are the same.  We have investigated this, and the discrepancy can be reduced (in many cases eliminated entirely) by projecting out the most \textit{stable} eigenmodes from the transient-growth calculations.  The methodology for doing this is described below.

First, in more detail, the evolution operator $\evolop_t=V\mathe^{\Lambda t}V^{-1}$ involves all eigenmodes that arise in the truncated Chebyshev expansion of the full problem.  It is well known that only the most unstable eigenmodes are computed accurately in the Chebyshev collocation method.  Therefore, contributions to the evolution operator coming from highly \textit{stable} modes will lead to numerical error.  Although such stable modes are of no interest an an eigemode analysis (or more generally, in any kind of analysis wherein the limit $t\rightarrow\infty$ is taken), they could interfere with transient-growth calculations at finite time.  Therefore, as in the standard literature on transient growth, the proposal here is to project such modes out of the evolution operator.

To do this, an approximate solution to the initial value problem~\eqref{eq:os_ivp} is proposed involving only the first $Q$ modes:
\begin{equation}
\mychi(t)=\sum_{q=1}^Q \kappa_q(t)\mychi_{(q)},\qquad \mychi_{(q)}=\left(w_{L(q)},w_{G(q)},\eta_{L(q)},\eta_{G(q)}\right)^T,\qquad Q\leq N,
\label{eq:qtrunc}
\end{equation}
where $L$ and $G$ label the phases.
The solution~\eqref{eq:qtrunc} is substituted into our calculation for the energy norm, which for a two-phase flow experiencing surface tension but no gravity, reads
\begin{equation}
E(t)=\frac{1}{2k^2}\bigg\{\left[\sum_{L,G}\int \left(|\partial_z w|^2+k^2|w|^2+|\eta|^2\right)\mathd z\right]+\frac{k^4}{\mywe}|\myfs|^2\bigg\},
\label{eq:et_twophase}
\end{equation}
where $We$ is the Weber number and $f_0$ represents the interface disturbance.  In order to do this calculation, we need for example expressions such as the following, valid for either one of the phases:
\[
\int |w|^2\mathd z=\int \left|\sum_q \kappa_q w_q(z)\right|^2\mathd z,
\]
which can be done in the Chebyshev expansion as follows:
\begin{eqnarray*}
\int |w|^2\mathd z&=&\int \left|\sum_q \kappa_q w_q(z)\right|^2\mathd z,\\
&=&\int \left|\sum_q \kappa_q \sum_{j}A_{j(q)}T_j(x)\right|^2\mathd z,\\
&=&\sum_q\sum_{q'}\kappa_q\kappa_{q'}^*\sum_j\sum_{j'}A_{j(q)}\mathbb{T}_{jj'}^{(0)}A_{j'(q')}^*.
\end{eqnarray*}
Similarly,
\[
\int |\eta|^2\mathd z
=\sum_q\sum_{q'}\kappa_q\kappa_{q'}^*\sum_j\sum_{j'}B_{j(q)}\mathbb{T}_{jj'}^{(0)}B_{j'(q')}^*,
\]
for either phase.
In this way, it is clear that
\begin{multline}
E(t)=\frac{1}{2k^2}\sum_{q,q'}\kappa_q\kappa_{q'}\left[\frac{k^4}{We}f_{0(q)}f_{0(q')}^*\right]\\
+
\frac{1}{2k^2}\sum_{q,q'}\kappa_q\kappa_{q'}\left[\sum_{L,G}\left(\sum_{jj'}A_{j(q)}\mathbb{T}_{jj'}^{(1)}A_{j'(q')}^*
+k^2\sum_{jj'}A_{j(q)}\mathbb{T}_{jj'}^{(0)}A_{j'(q')}^*
+\sum_{jj'}B_{j(q)}\mathbb{T}_{jj'}^{(0)}B_{j'(q')}^*\right)
\right].
\label{eq:et_twophase1}
\end{multline}
Clearly, this is a grotesque expression but it can be ameliorated.  Let us momentarily revert to an eigenvalue with a single component, with $[V,D]=\mathrm{eig}\left(\mathcal{L},\mathcal{M}\right)$ say, where the $q^\mathrm{th}$ column of the matrix $V$ corresponds to the $q^{\mathrm{th}}$ eigenvector, with $V_{(q)}=(A_{0(q)},\cdots,A_{N(q)})^T$ (say).  Then, for any $N\times N$ {\textit{symmetric}} matrix $C$, we have
\begin{eqnarray}
(V^\dagger C V)_{q'q}&=&\sum_{jj'}V^{\dagger}_{q'j'}C_{j'j}V_{jq},\nonumber\\
&=&\sum_{jj'}V^{*}_{j'q'}C_{j'j}V_{jq},\nonumber\\
&=&\sum_{jj'}A_{j'(q')}^*C_{j'j}A_{j(q)},\nonumber\\
&=&\sum_{jj'}A_{j'(q')}^*C_{jj'}A_{j(q)}\label{eq:resmx}
\end{eqnarray}
Hence, by introducing the matrix
\[
\mathbb{T}=\left(\begin{array}{ccccc}\mathbb{T}^{(1)}_L+k^2\mathbb{T}^{(0)}_L& & & & \\
 & \mathbb{T}^{(1)}_G+k^2\mathbb{T}^{(0)}_G & & &\\
& & \mathbb{T}^{(0)}_L & & \\
& & & \mathbb{T}^{(0)}_G & \\
& & & & k^4/We \end{array}\right)
\]
it should be clear from Equation~\eqref{eq:resmx} that Equation~\eqref{eq:et_twophase} can be rewritten as
\begin{equation}
E(t)=\frac{1}{2k^2}\sum_{qq'}\kappa_q\kappa_{q'}^* \left(V_Q^\dagger\mathbb{T}V_Q\right)_{q'q},
\label{eq:et_twophase2}
\end{equation}
where now $V_Q$ is an $N\times Q$ matrix where each column is an eigenvector of the full problem, where $N=2(N_L+N_G+2)+1$; in other words, the $q^{\mathrm{th}}$ column of $V_Q$ is the eigenvector
\[
V_{(q)}=
(A_{0(q)}^L,\cdots,A_{N_L(q)}^L,
 A_{0(q)}^G,\cdots,A_{N_G(q)}^G,
 B_{0(q)}^L,\cdots,B_{N_L(q)}^L,
 B_{0(q)}^G,\cdots,B_{N_G(q)}^G
)^T.
\]
We consider again Equation~\eqref{eq:et_twophase2}.  Because the $\kappa_q$'s are weights in an eigenfunction expansion of the solution of a linear evolutionary equation, we have $\kappa_q=\mathe^{\lambda_q t}\mu_q$, where $\mu_q$ is a constant.  Thus,
\begin{eqnarray*}
E(t)&=&\frac{1}{2k^2}\sum_{qq'}\kappa_q\kappa_{q'}^* \left(V_Q^\dagger\mathbb{T}V_Q\right)_{q'q},\\
&=&\frac{1}{2k^2}\sum_{qq'}\mu_{q'}^* \mathe^{\lambda_{q'}^*t} \left(V_Q^\dagger\mathbb{T}V_Q\right)_{q'q}\mathe^{\lambda_qt}\mu_q,\\
&=&\frac{1}{2k^2}\sum_{qq'}\mu_{q'}^* \left(\mathe^{\Lambda^*t}V_Q^\dagger\mathbb{T}V_Q\mathe^{\Lambda t}\right)_{q'q}\mu_q,\\
&:=&\frac{1}{2k^2}\langle\vec{\mu},\left(\mathe^{\Lambda^*t}V_Q^\dagger\mathbb{T}V_Q\mathe^{\Lambda t}\right)\vec{\mu}\rangle,
\end{eqnarray*}
where the last line makes use of an obvious notation.  The relation
\begin{equation}
E_Q(t)=\frac{1}{2k^2}\langle\vec{\mu},\left(\mathe^{\Lambda_Q^*t}V_Q^\dagger\mathbb{T}V_Q\mathe^{\Lambda_Q t}\right)\vec{\mu}\rangle
\label{eq:et_twophaseQ}
\end{equation}
 is our final expression for the energy of a disturbance involving only the first $Q$ eigenmodes: a subscript $Q$ has been introduced (with $E(t)\rightarrow E_Q(t)$) to remind ourselves that only $Q$ eigenmodes are used in the expansions.

We carry out several consistency checks on our calculations.  The first one involves checking that the matrix multiplication $V_Q^{\dagger}\mathbb{T}V_Q$ makes sense.  We first of all note the size of the various matrices: $V_Q$ is a $N\times Q$ matrix and $\mathbb{T}$ is a matrix of size $N\times N$.  Hence, doing the `cross multiplication' check to see if the product $V_Q^{\dagger}\mathbb{T}V_Q$ is defined, we have
\[
(Q\times N)\times (N\times N)\times (N\times Q)=(Q\times Q),
\]
and the matrix multiplication is therefore consistent.  Next, we would also like to check that $E_{Q=N}(t)$ agrees with our earlier expressions for the energy, recalled here from Section~\ref{sec:basic} as
\begin{eqnarray}
E(t)&=&\frac{1}{2k^2}\langle \vec{x}, \evolop^\dagger \mathbb{T}\evolop\vec{x}\rangle,\nonumber\\
&=& \frac{1}{2k^2}\langle\vec{x}, V^{-1\dagger}\mathe^{\Lambda^*t}V^\dagger\mathbb{T}V\mathe^{\Lambda t}V^{-1}\vec{x}\rangle,\nonumber\\
&=& \frac{1}{2k^2}\langle V^{-1}\vec{x}, \left(\mathe^{\Lambda^*t}V^\dagger\mathbb{T}V\mathe^{\Lambda t}\right)\left(V^{-1}\vec{x}\right)\rangle,\label{eq:check}
\end{eqnarray}
Clearly, if we set $\vec{\mu}=V^{-1}\vec{x}$ in Equation~\eqref{eq:check} and $Q=N$ in Equation~\eqref{eq:et_twophaseQ} we have
\[
E_{N=Q}=E(t)=\frac{1}{2k^2}
\langle V^{-1}\vec{x}, \left(\mathe^{\Lambda^*t}V^\dagger\mathbb{T}V\mathe^{\Lambda t}\right)\left(V^{-1}\vec{x}\right)
=\frac{1}{2k^2}
\langle \vec{\mu}, \left(\mathe^{\Lambda^*t}V^\dagger\mathbb{T}V\mathe^{\Lambda t}\right)\vec{\mu}\rangle,
\]
and the two approaches are identical (and hence consistent) when $N=Q$.  However, even when $Q<N$, we can identify $\vec{x}=V_Q\vec\mu$ in Equation~\eqref{eq:et_twophaseQ}, which will be helpful in what follows.

We obtain the maximum growth $G(t)$ in the usual way by maximizing the constrained problem
\[
E_Q[\vec{\mu}]=\frac{1}{2k^2}
\langle\vec{\mu},\left(\mathe^{\Lambda_Q^*t}V_Q^\dagger\mathbb{T}V_Q\mathe^{\Lambda_Q t}\right)\vec{\mu}\rangle
-\lambda\left[\frac{1}{2k^2}
\langle\vec{\mu},\left(V_Q^\dagger\mathbb{T}V_Q\right)\vec{\mu}\rangle-1\right],
\]
As before, the maximum growth rate is obtained by setting
\[
\frac{\delta E_Q}{\delta\vec{\mu}^*}=0,
\]
hence
\[
\left(\mathe^{\Lambda_Q^*t}V_Q^\dagger\mathbb{T}V_Q\mathe^{\Lambda_Q t}\right)\vec{\mu}
=\lambda \left(V_Q^\dagger\mathbb{T}V_Q\right)\vec{\mu}
\]
and hence
\[
\sup_{\vec{\mu}}E_Q[\vec{\mu}]=\max \lambda=G(t).
\]
The optimal vector is the corresponding eigenvector of the problem.  In the usual basis, the optimal vector is given by the transformation $\vec{x}=V_Q\vec{\mu}$.

\subsection{Validation -- two-phase flow}

We have validated these calculations with respect to Figures~4 and~5 in Reference~\cite{yecko2008disturbance} and the results are shown below in Figure~\ref{fig:xcompare4} and~\ref{fig:xcompare5}.  In our calculations, we used $n_1=n_{2}=42$ Chebyshev collocation points in each phase, and carried out the transient-growth calculations using only the first 10 eigenmodes.  These choices were made on a trial-and-error basis: $(n_1,n_2)$ were varied to obtain numerical convergence, and the cutoff of number of eigenmodes was varied so as to find maximum agreement between the present calculations and those in the reference text.
There are some discrepancies between the two approaches in Figure~\ref{fig:xcompare4} while no such discrepancies are in evidence in Figure~\ref{fig:xcompare5}.  The small discrepancies are not concerning, since the results depend slightly on the cutoff number of eigenmodes in the calculations, and this is not stated explicitly in the reference text.  However, the excellent agreement between the two approaches in Figure~\ref{fig:xcompare5} reinforces our confidence in our own implementation of the transient-growth calculation.
\begin{figure}[htb]
	\centering
		\includegraphics[width=0.45\textwidth]{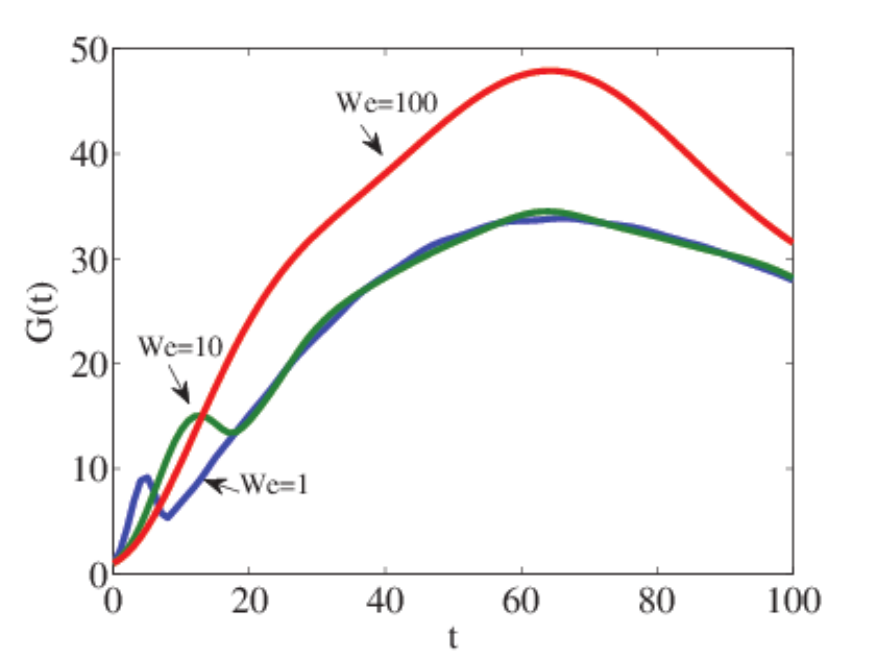}
		\includegraphics[width=0.45\textwidth]{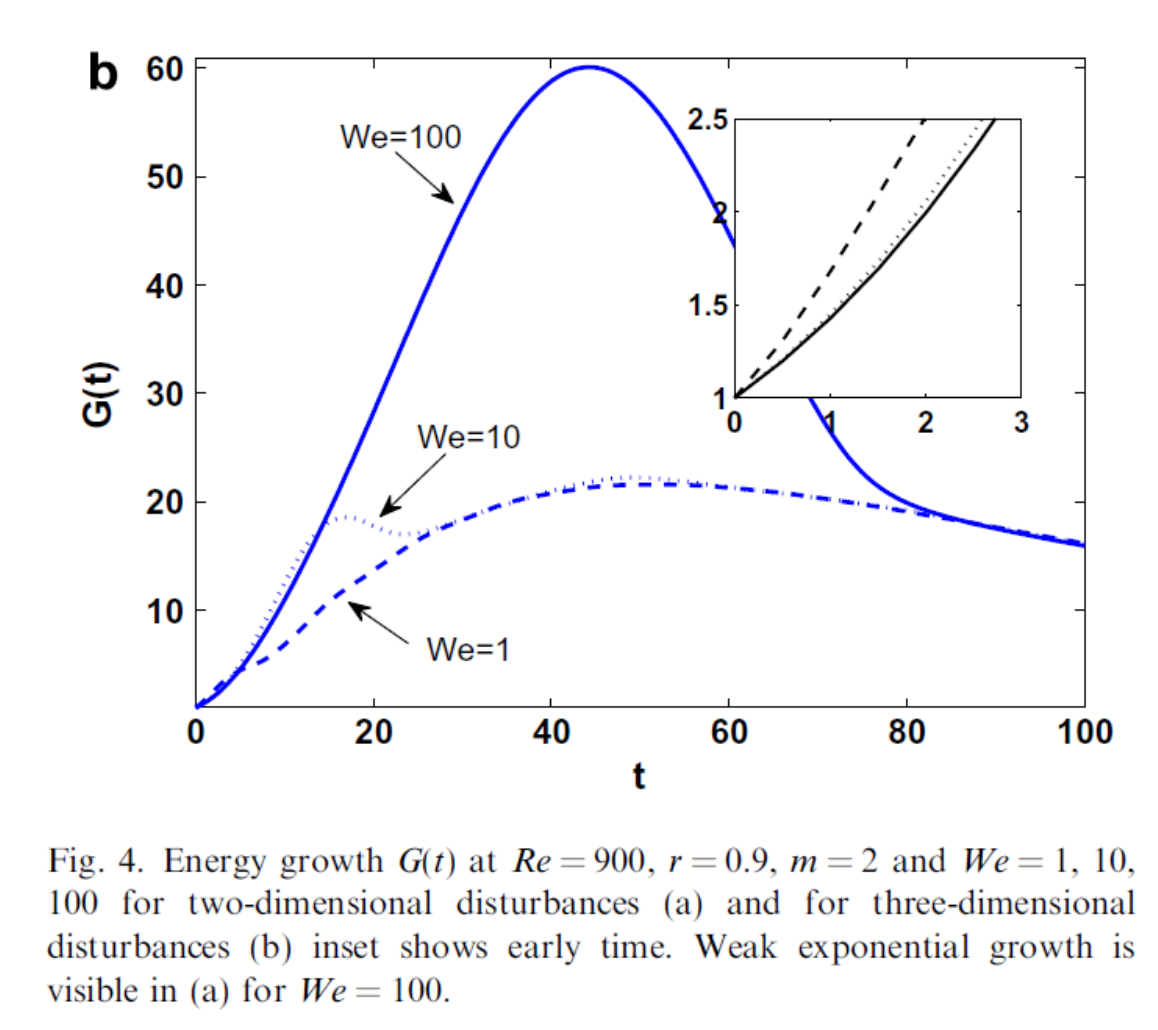}
		\caption{Comparison with Figure~4 in Reference~\cite{yecko2008disturbance}: panel (a) this work; panel (b): taken directly from the reference text}
	\label{fig:xcompare4}
\end{figure}
\begin{figure}[htb]
	\centering
		\includegraphics[width=0.45\textwidth]{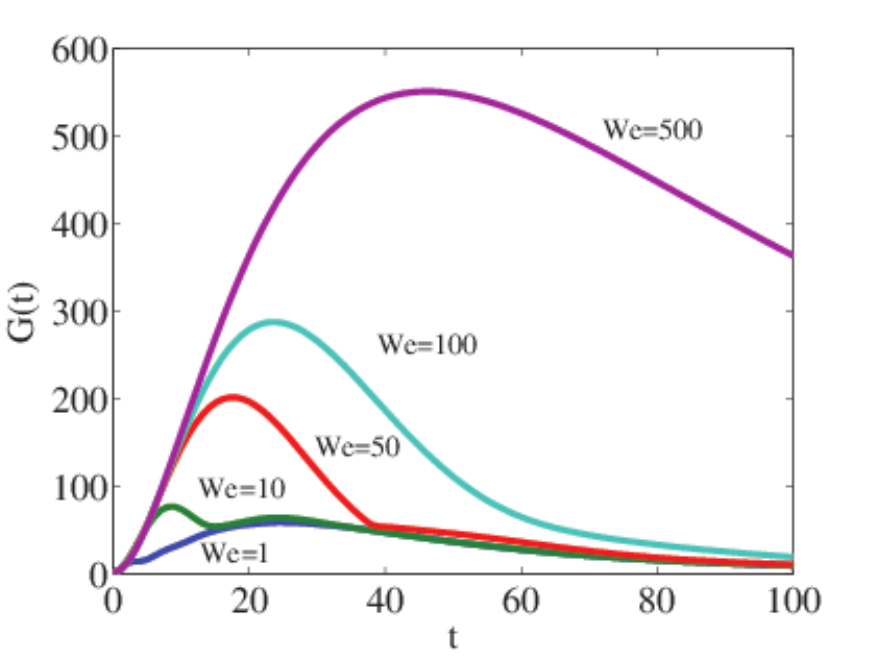}
		\includegraphics[width=0.45\textwidth]{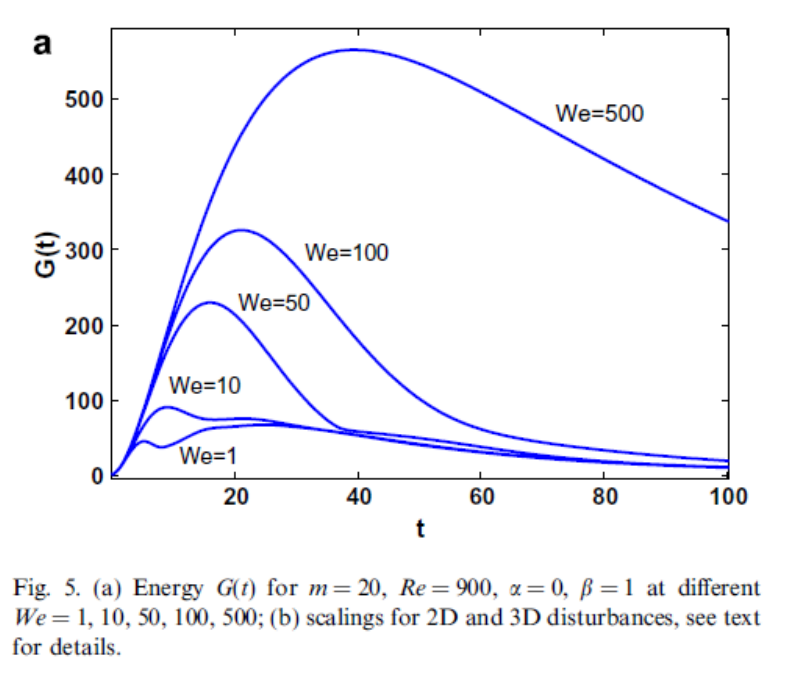}
\caption{Comparison with Figure~5 in Reference~\cite{yecko2008disturbance}: panel (a) this work; panel (b): taken directly from the reference text}
	\label{fig:xcompare5}
\end{figure}

\subsection{Linearized direct numerical simulation}

We demonstrate how a linearized version of the two-phase Navier--Stokes equations can be solved numerically.  We start with the already-linearized Orr--Sommerfeld--Squire equations in Equation~\eqref{eq:oss_laplace}.  Reversing the Laplace transform implicit  in these equations, we obtain the following Cauchy problem:
\begin{equation}
\frac{\partial}{\partial t}\left(\begin{array}{cc}\mathcal{M}_{OS} & 0\\
                       \mathcal{M}_{C} & \mathcal{M}_{S}\end{array}\right)\left(\begin{array}{c}\wzonlyab(z,t)\\\omegazonlyab(z,t)\end{array}\right)
=
\left(\begin{array}{cc}\mathcal{L}_{OS} & 0\\
                       \mathcal{L}_{C} & \mathcal{L}_{S}\end{array}\right)\left(\begin{array}{c}\wzonlyab(z,t)\\\omegazonlyab(z,\lambda)\end{array}\right),\qquad t>0,
\label{eq:cauchy}
\end{equation}
subject to arbitrary initial data 
\begin{equation}
\wzonlyab(z,t=0)=w_0(z),\qquad \omegazonlyab(z,t=0)=\eta_0(z).
\label{eq:cauchy_init}
\end{equation}
An approximate solution to the initial value problem~\eqref{eq:cauchy} is proposed involving only the first $Q$ eigenmodes modes of the Orr--Sommerfeld--Squire (OSS) theory:
\begin{equation}
\left(\begin{array}{c}\wzonlyab(z,t)\\\omegazonlyab(z,\lambda)\end{array}\right)\approx \sum_{q=1}^Q \kappa_q(t)\bm{\chi}_{q}(z,t),
\label{eq:qtrunc}
\end{equation}
where $\bm{\chi}_{q}(z,t)=\left(\wzonlyab^{(q)}(z,t),\omegazonlyab^{(q)}(z,\lambda)\right)^T$ denotes the $q^{\mathrm{th}}$ OSS eigenmode, ordered in terms of the eigenvalues with the largest real part, with $q=1$ corresponding to the most-dangerous mode.  It is straightforward to see that $\kappa_q(t)=a_q\mathe^{\lambda_q t}$, where $a_q$ is a constant and $\lambda_q$ is the $q^{\mathrm{th}}$ eigenvalue of the OSS theory.  A linear transformation then gives the $a_q$'s in terms of the initial data~\eqref{eq:cauchy_init}, as already described in the above. 

\subsection{The resolvent}

The resolvent of the evolutionary equation in the energy norm can also be computed along these lines.  By definition,
\begin{equation}
R(z)=\|\left(\mathcal{L}_{\bm{\alpha}}-z\mathcal{M}_{\bm{\alpha}}\right)^{-1}\|_E,
\label{eq:Rz}
\end{equation}
with $R(z)=\infty$ when $z\in\mathrm{spec}(\mathcal{L}_{\bm{\alpha}},\mathcal{M}_{\bm{\alpha}})$.  Introduce $T=\mathbb{T}/2k^2$ and $\resmx=\left(\mathcal{L}_{\bm{\alpha}}-z\mathcal{M}_{\bm{\alpha}}\right)^{-1}$.  Using the definition~\eqref{eq:Rz} and the definition of the operator norm, it should be clear that Equation~\eqref{eq:Rz} can be rewritten as
\begin{equation}
[R(z)]^2=\sup_{\langle\vecx,T\vecx\rangle=1}\langle \resmx\vecx,T\resmx\vecx\rangle,
\label{eq:Rz_opt}
\end{equation}
where the angle brackets denote the usual scalar product.  We introduce a constrained functional,
\[
F[\vecx]=\langle \resmx\vecx,T\resmx\vecx\rangle-\lambda\left(\langle \vecx,T\vecx\rangle-1\right),
\]
such that $\delta F/\delta\vecx^*=0$ solves the optimization problem~\eqref{eq:Rz_opt}, hence
\begin{equation}
\resmx^\dagger T \resmx\vecx=\lambda T\vecx,
\label{eq:gevp_res}
\end{equation}
hence
\[
[R(z)]^2=\max\big\{\mathrm{spec}\left[\resmx^\dagger T \resmx,T\right]\big\},
\]
where $\mathrm{spec}\left[\resmx^\dagger T \resmx,T\right]$ denotes the spectrum of the generalized eigenvalue problem~\eqref{eq:gevp_res}. 

\section{Direct Numerical Simulation -- Illustrative example}

We consider here first a test case to assess whether the early-stage flow behaviour is primarily through modal or transient growth, using the linearized and the full DNS. This also serves to validate the full DNS.
The following parameter values are used: $m=50,h_0=0.2,\mygrav=0.1,\mywe=10)$, with $r=1000$ and $\myre=500$. 
The initial configuration is a hydrostatic pressure field, with $\vecu=0$. We first choose an initial interfacial disturbance that is simply a superposition of Fourier modes that fit in the domain and have equal amplitude,
\begin{equation}
\myfs(x,y,t=0)=h_0+\tfrac{1}{16}A_0\sum_{m=0}^3\sum_{n=0}^3\cos\left[(2\pi/L_x)mx+(2\pi/L_y)ny+\varphi_{mn}\right],
\label{eq:ic1}
\end{equation}
where $A_0$ is the initial amplitude of the disturbance and $\varphi_{mn}\in[0,2\pi)$ is a random phase. We return to the choice of initial condition below.
For the numerical simulation, a uniform grid is used and the number of grid points in the $z$-direction is $340$.  The length of the channel corresponds to the most-dangerous mode from linear theory and the width is $L_y=1$.  The timestep is taken to be $10^{-5}$.  These numerical parameters are found to be sufficient to ensure convergence of the numerical method.

\begin{figure}
	\centering
		\subfigure[$\,\,\alpha=\alpha_0,\beta=0$]{\includegraphics[width=0.32\textwidth]{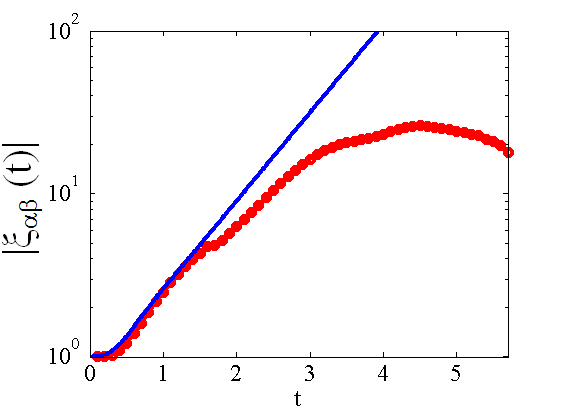}}
		\subfigure[$\,\,\alpha=2\alpha_0,\beta=0$]{\includegraphics[width=0.32\textwidth]{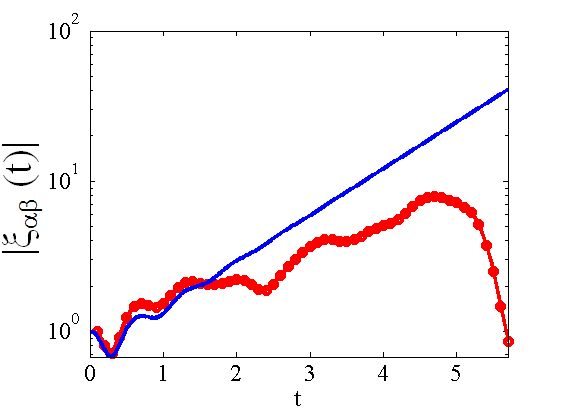}}\\
		\subfigure[$\,\,\alpha=0,\beta=\beta_0$]{\includegraphics[width=0.32\textwidth]{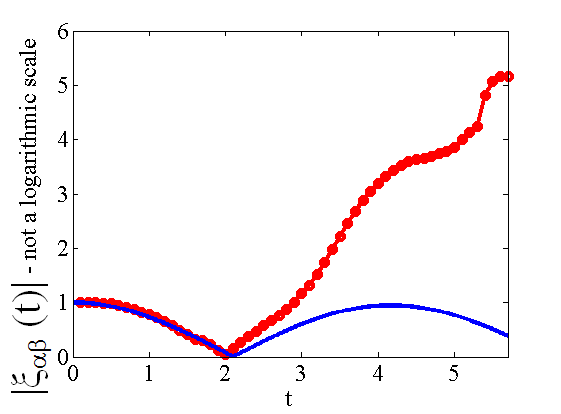}}
		\subfigure[$\,\,\alpha=\alpha_0,\beta=\beta_0$]{\includegraphics[width=0.32\textwidth]{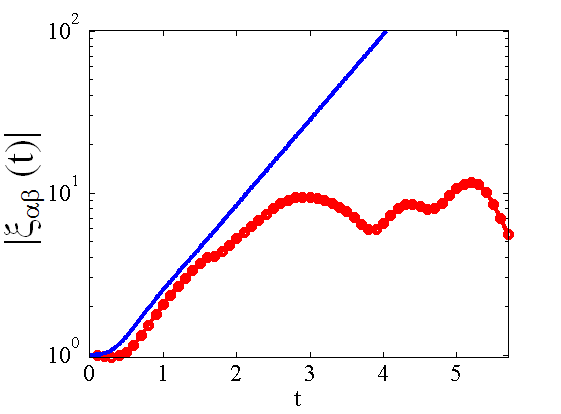}}
		\subfigure[$\,\,\alpha=2\alpha_0,\beta=\beta_0$]{\includegraphics[width=0.32\textwidth]{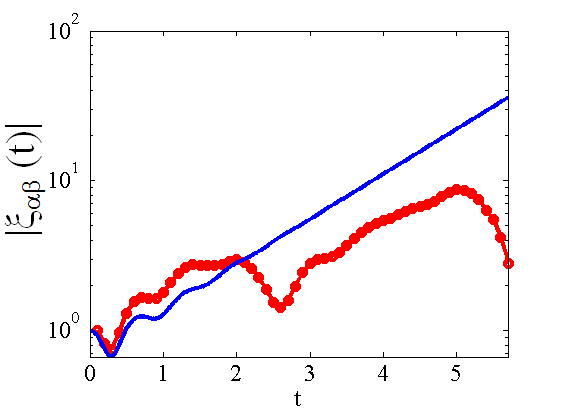}}
		\caption{DNS results (lines with circles) for the  case $(m=50,h_0=0.2,\mygrav=0.1,\mywe=10)$, with $r=1000$ and $\myre=500$. Shown also is a comparison with linearized DNS (unadorned lines).  Here, $\alpha_0=2\pi/L_x$ and $\beta_0=2\pi/L_y$ denote the fundamental wavenumber in the streamwise and spanwise directions respectively. In panel (c) the growth of the relevant amplitude is modest and a vertical linear (as opposed to logarithmic) scale is used.  Also, the `kink' at $t=2$ in the same panel simply corresponds to a change in sign of the $\myfs_{\alpha\beta}(t)$.}
	\label{fig:highRe_spectra}
\end{figure}

The analysis of the subsequent wave evolution is based on interfacial spectra, defined as follows:
the interfacial elevation $\myfs(x,y,t)$ (where single-valued) is extracted from the levelset function $\phi(x,y,z,t)$ and a Fourier analysis is conducted, such that the interfacial spectra $\myfs_{\alpha\beta}(t)$ are obtained at each timestep, where
\begin{equation}
\myfs_{\alpha\beta}(t)=\int_{0}^{L_x}\mathd x\int_0^{L_y}\mathd y\,\myfs(x,y,t)\mathe^{-\imag\alpha x-\imag\beta y}.
\label{eq:halphabetadef}
\end{equation}
Results based on the interfacial spectra~\eqref{eq:halphabetadef} from the full DNS are shown in Figure~\ref{fig:highRe_spectra}, alongside a comparison with linear theory.  The results from the linear theory are obtained by linearized DNS, which in turn is based on solving the Orr--Sommerfeld--Squire equation as an initial-value problem, as opposed to an eigenvalue problem.
The figure indicates good agreement between the two independent methods, at least until the onset of wave overturning at $t\approx 2$.  

For very early times  in Figure~\ref{fig:highRe_spectra}, the amplitudes of the different modes behave in a variety of different ways and no clear-cut behaviour is observed.  This is because the initial condition comprises a zero perturbation velocity field and a sinusuoidal interface disturbance.  As such, for each wavenumber $(\alpha,\beta)$, these initial conditions correspond to a superposition of Orr--Sommerfeld--Squire modes, each varying in time by a distinct complex exponential frequency.  As time goes by, the most-dangerous Orr--Sommerfeld--Squire mode at a particular wavenumber is selected, leading to clearly visible
%
%it can be seen that 
exponential growth of the amplitudes  for $t\apprge 0.5$, corresponding to the corresponding modal growth rate.
The growth rate of streamwise and spanwise modes are equally prominent, as the modal growth rate $\omega_{\imag}(\alpha,\beta)$ falls off only very slowly with increasing $\beta$. Thus, these results indicate that a direct route to linear instability (and subsequent nonlinear instability) pertains at high density ratios, owing to the particular structure of the growth rate $\omega_{\mathrm{i}}(\alpha,\beta)$ of the linearized dynamics.

\begin{figure}[htb]
	\centering
	  \subfigure[$\,\,r=1\,\, (t=0.01)$]   {\includegraphics[width=0.235\textwidth]{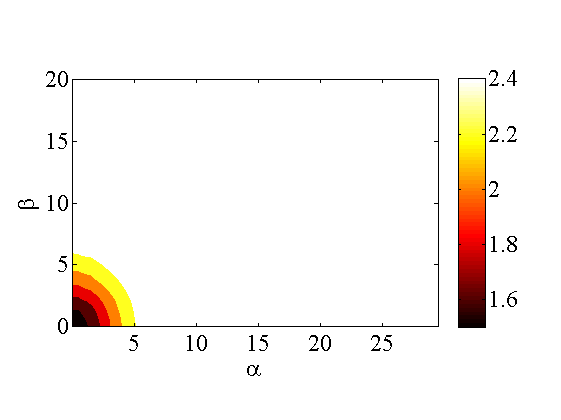}}
		\subfigure[$\,\,r=1\,\, (t=0.05)$]   {\includegraphics[width=0.235\textwidth]{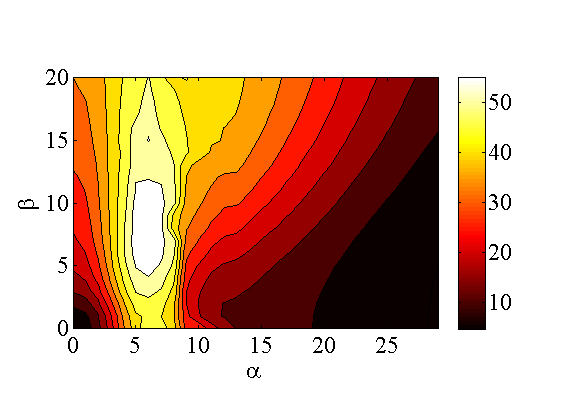}}
		\subfigure[$\,\,r=1\,\, (t=0.1)$]{\includegraphics[width=0.235\textwidth]{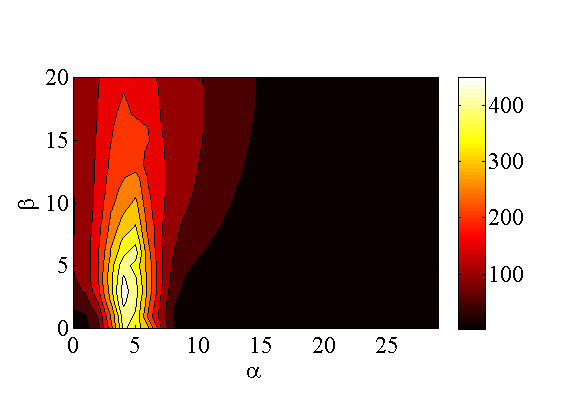}}
		\subfigure[$\,\,r=1\,\, (t=0.2)$]{\includegraphics[width=0.235\textwidth]{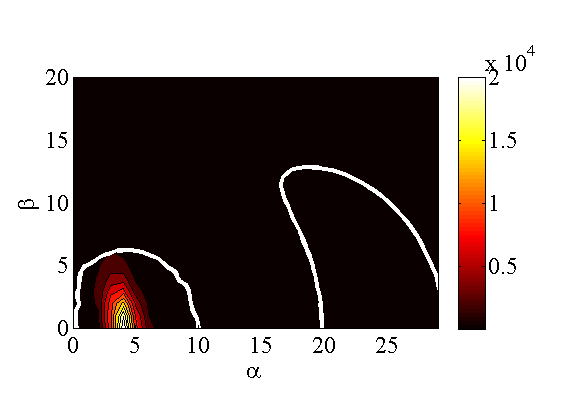}}\\
    \subfigure[$\,\,r=1000\,\, (t=0.05)$]   {\includegraphics[width=0.23\textwidth]{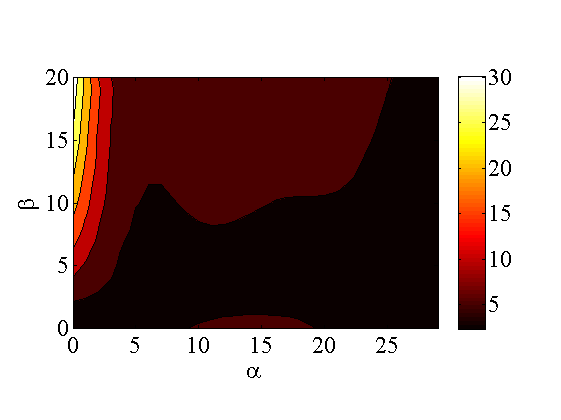}}
		\subfigure[$\,\,r=1000\,\, (t=0.1)$]{\includegraphics[width=0.235\textwidth]{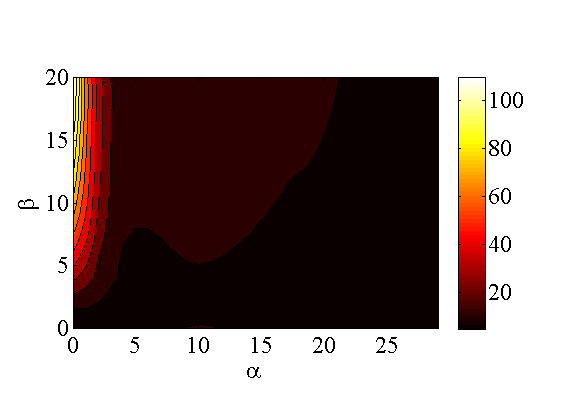}}
		\subfigure[$\,\,r=1000\,\, (t=0.5)$]{\includegraphics[width=0.235\textwidth]{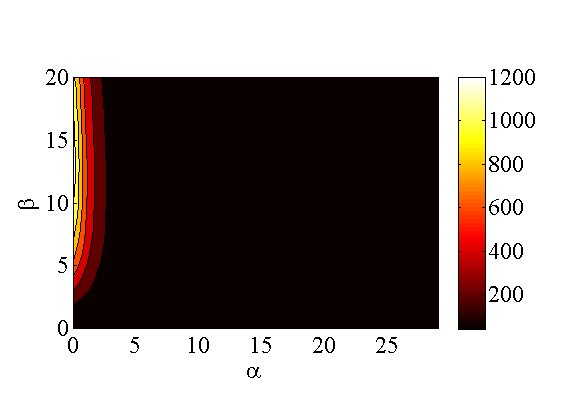}}
		\subfigure[$\,\,r=1000\,\, (t=1.25,\,t=2.5)$]{\includegraphics[width=0.235\textwidth]{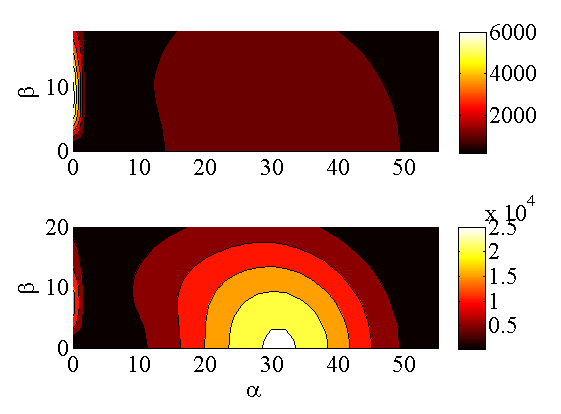}}
	\caption{Evolution of linearized DNS showing the maximum amplification factor $G_{\alpha\beta}(t)$ as a function of $t$.  
	Parameters: $(m=50,h_0=0.2,\mygrav=0.1,\mywe=10)$, and $\myre=500$.
		Across the top: $r=1$.  Across the bottom: $r=1000$.   The white curves in panel (d) denote the neutral curves for $r=1$ where $\omi(\alpha,\beta)=0$ in the eigenvalue linear stability analysis.  The second curve at large $\alpha$-values corresponds to the second mode identified in the main text.   The corresponding neutral curve for $r=1000$ is located outside the wavenumber range shown in the figure.
		}
	\label{fig:trans_r1}
\end{figure}

The extremely large spanwise amplification factors expected from trasneint-growth theory do not manifest themselves in  Figure~\ref{fig:highRe_spectra} (neither in the full nor in the linearized DNS); no transient growth is observed. 
The reason is that such transient amplification factors (e.g. Figure~\ref{fig:trans_r1}) are the
\textit{envelope} corresponding to the maximum realizable growth corresponding to a  particular initial condition (i.e. the one that optimizes the energy norm at a particular time).

Although transient growth is not observed in the simulations presented above and in subsequent sections, it is possible to choose highly specialized initial conditions, that are suggested by the linear transient growth theory, that do lead to non-modal growth. We have therefore performed further direct numerical simulations for the case  $(r,m,\myre,h_0,\mygrav,\mywe)=(10,50,500,0.2,0.1,10)$, choosing as the initial conditions on the basis of Figure~\ref{fig:trans_r1}, which suggests that for the present parameter choice, transient growth of spanwise-only modes is favoured at early times up to $t\approx 1$.  That is,  the initial condition is precisely those values that maximize the energy norm at $t=0.1$ in the linear theory, for which we can expect very strong amplification of the mode $(\alpha,\beta)=(0,2\pi/L_y)$ (the amplitude of the disturbances is selected such that the initial amplitude of the sinusoidal wave on the interface is $0.0017$).
\begin{figure}
	\centering
		\includegraphics[width=0.6\textwidth]{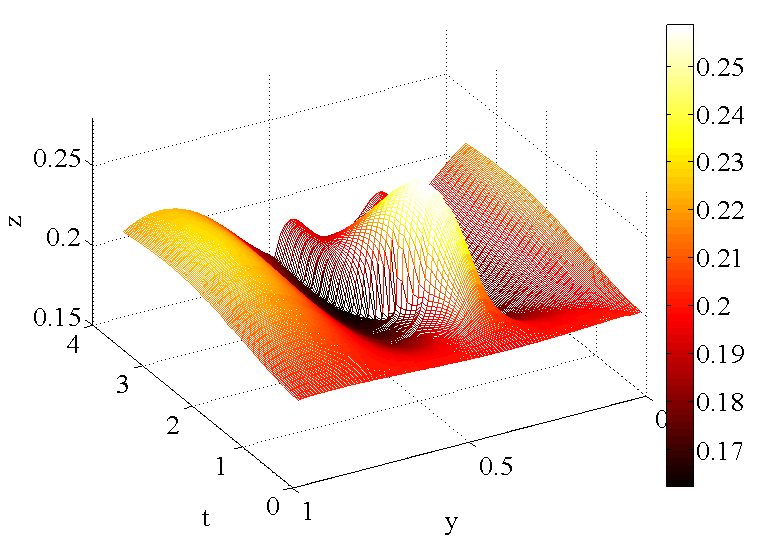}
\caption{
DNS results for the  case $(m=50,h_0=0.2,\mygrav=0.1,\mywe=10)$, with $r=1000$ and $\myre=500$. Shown is a snapshot of a slice ($yz$-plane) of the interfacial configuration at various times, for an initial condition chosen so as to maximize the energy norm of the mode $(\alpha,\beta)=(0,2\pi/L_y)$ in the linear theory at $t=0.1$.
}
	\label{fig:hxy1_t}
\end{figure}
A slice of the resulting interfacial configuration is shown in Figure~\ref{fig:hxy1_t} at various times, showing the variation of the interface in the spanwise direction (no streamwise variations are detected in the course of the simulation).    The corresponding velocity field at $t=2$ is shown in Figure~\ref{fig:uvw_200000}. 
\begin{figure}
	\centering
		\subfigure[$\,\,u$]{\includegraphics[width=0.3\textwidth]{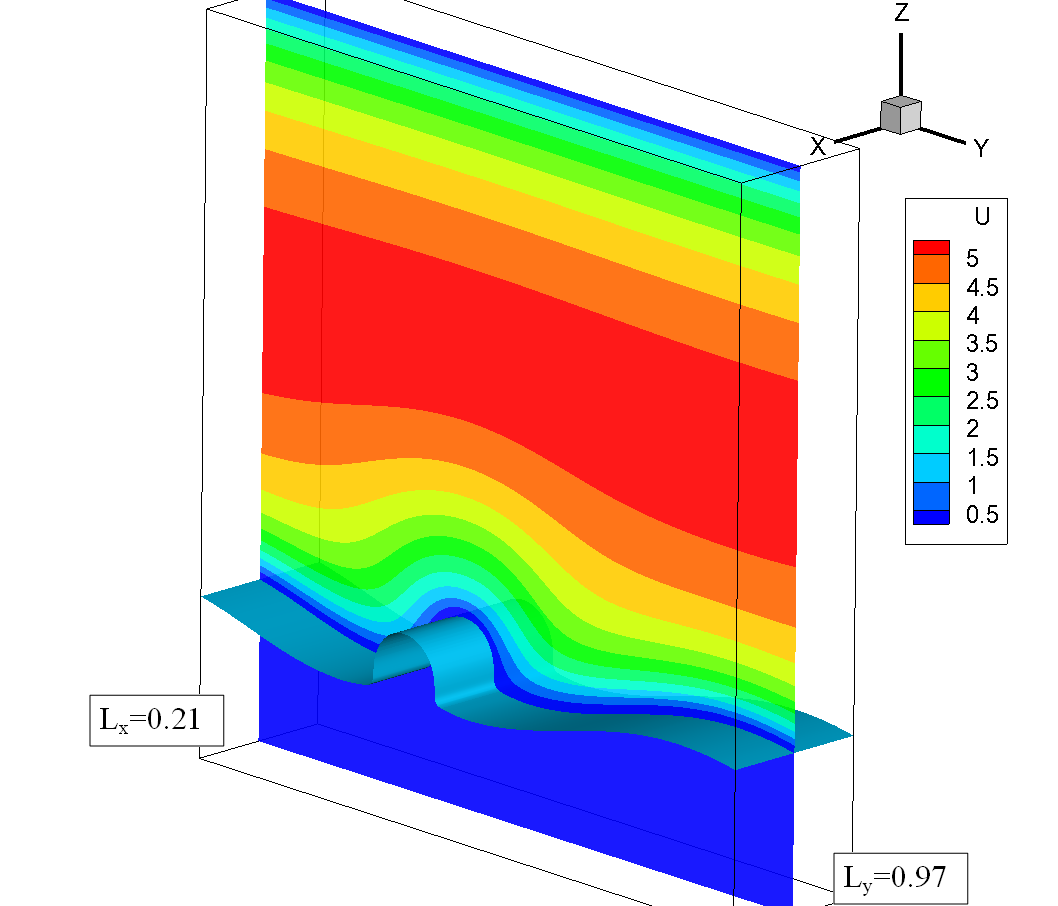}}
		\subfigure[$\,\,v$]{\includegraphics[width=0.3\textwidth]{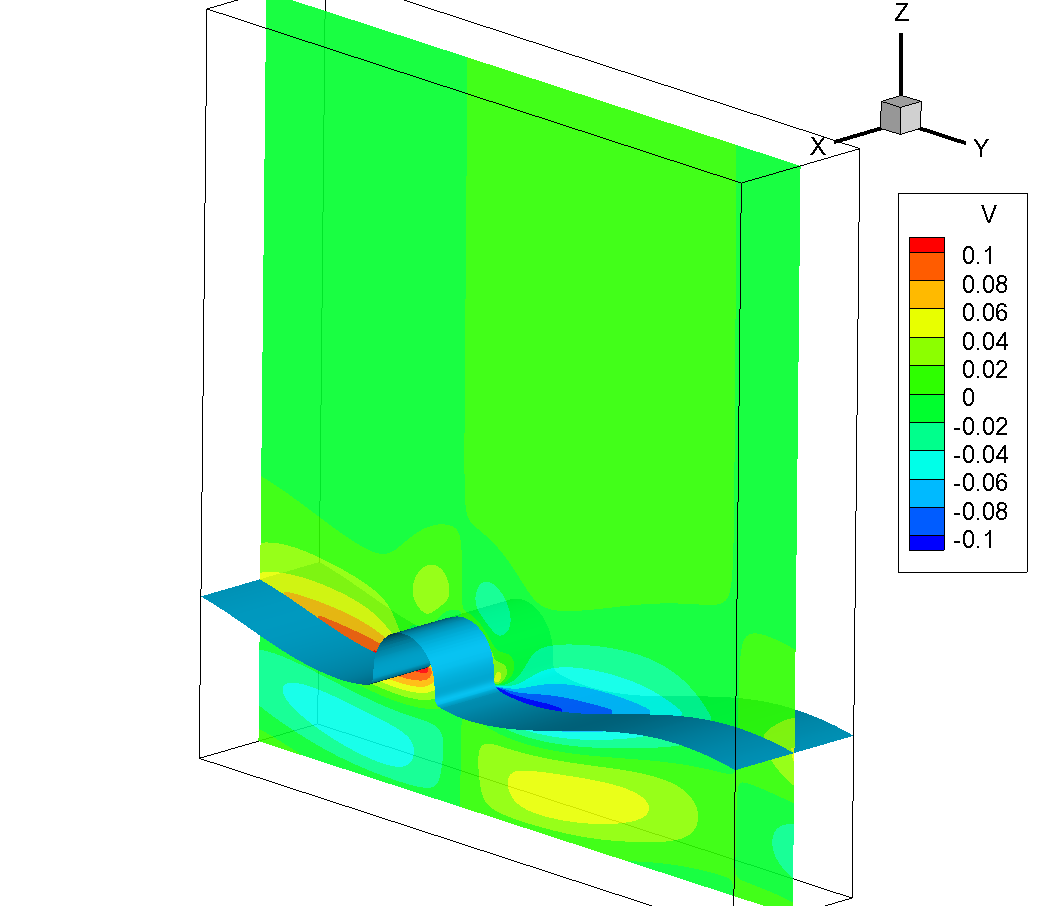}}
		\subfigure[$\,\,w$]{\includegraphics[width=0.3\textwidth]{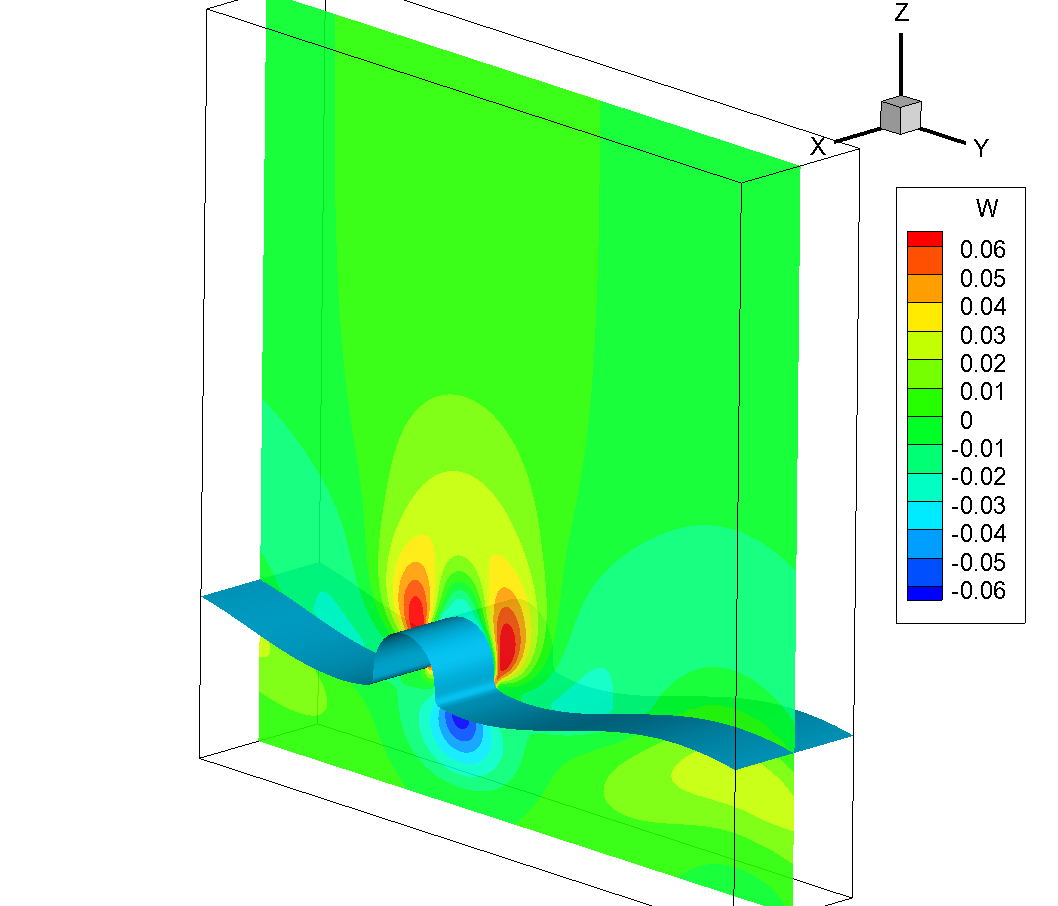}}
		\caption{Velocity fields at $t=2$ for the case $(m=50,h_0=0.2,\mygrav=0.1,\mywe=10)$, with $r=1000$ and $\myre=500$.  The initial conditions are those of the transient-growth study.}
	\label{fig:uvw_200000}
\end{figure}
It can be seen that the system responds strongly to the chosen initial condition and spanwise interfacial wave grows rapidly in amplitude before decaying again at later times. 
This is fully consistent with the linear theory of transient growth, in spite of the finite wave amplitudes in the simulation.  Moreover, the observed late-time decay of the wave amplitude is consistent with linear theory; this decay occurs because the eigenvalue theory (valid at late times) predicts that the mode $(\alpha,\beta)=(0,2\pi/L_y)$ is linearly stable.  The conclusion here therefore is that transient growth fails to trigger a self-sustained nonlinear disturbance~\cite{Schmid2001,Waleffe1998,naraigh2015} which could give rise to subsequent nonlinear dynamics.

\section{Conclusions}

Summarizing, we have introduced and validated an algorithm to compute transient amplification factors for the Orr--Sommerfeld--Squiare linear theory for parallel two-phase flow.  We have rigorously compared the predictions of the method with direct numerical simulation in a particular case -- chosen so as to be strongly supercritical, in the sense that the modal growth rates are large.   In this case, the modal growth rates dominate over the transient growth.  This points towards a general conclusion that when modal growth rates are strong, transient growth takes a back seat.

%\bibliographystyle{unsrt}
%\bibliography{turbulence_bibliography}

\end{document}